\documentclass[reqno]{amsart}

\usepackage[margin=3cm]{geometry}

\usepackage{arXiv_style}
\usepackage{fontawesome}

\makeatletter
\def\blfootnote{\gdef\@thefnmark{}\@footnotetext}
\makeatother

\title[On \texorpdfstring{$\gamma$}{gamma}-Contraction and \texorpdfstring{$\beta$}{beta}-Contraction]{\LARGE%
    On \texorpdfstring{$\gamma$}{gamma}-Contraction and \texorpdfstring{$\beta$}{beta}-Contraction%
    \\[.2em]%
    \normalsize%
    A Unified Framework for Colour-Preserving Graph Reduction%
}

\author[E.\ Onofri et al.]{}

\begin{document}

\blfootnote{$^{\star}$ (\href{mailto:elia.onofri@kaust.edu.sa}{\faEnvelopeO}) \texttt{elia.onofri@kaust.edu.sa}, (\href{https://www.eliaonofri.it}{\faGlobe}) \texttt{www.eliaonofri.it}}

\maketitle

\vspace{-0.5em}

\begin{center}
    \begin{minipage}{.89\linewidth}\centering
        \textsc{Elia Onofri}$^{\, \star,\, \orcidlink{0000-0001-8391-2563}}$.
        \\
        \bigskip
        \begin{minipage}{.9\linewidth}\centering
            \footnotesize
            Computer, Electrical and Mathematical Sciences and Engineering (CEMSE) Division,\\
            King Abdullah University of Science and Technology (KAUST)\\
            Thuwal 23955, Saudi Arabia
        \end{minipage}
    \end{minipage}
\end{center}

\medskip
\thispagestyle{empty}

\begin{abstract}

    Graphs are a fundamental abstraction in computer science and discrete mathematics, where information is encoded in their combinatorial structure.
    Graph-reduction techniques aim at simplifying graphs while preserving selected structural properties, typically by grouping vertices and replacing each group with a representative, yielding a contracted graph.
    
    \medskip\noindent
    A common instance of this paradigm arises when vertices carry categorical information, formalised as a colouring of the vertex set.
    In this setting, natural contraction units correspond to connected components of vertices sharing the same colour.
    
    \medskip\noindent
    In this work, we provide a rigorous mathematical formalisation of \emph{$\gamma$-contraction}, a colour-based graph contraction operation.
    We interpret $\gamma$-contraction as a quotient-like construction that preserves categorical connectivity, and clarify its relationship with classical notions of graph contraction and quotient graphs.
    To support a constructive and algorithmic treatment, we introduce a locally-defined iterative variant, termed \emph{$\beta$-contraction}, which captures the core mechanism underlying $\gamma$-contraction.
    
    \medskip\noindent
    Building on this framework, we analyse the contraction process from a theoretical perspective and establish formal guarantees of correctness and convergence.
    In particular, we prove that $\beta$-contraction converges in a logarithmic number of iterations to $\gamma$-contraction, and that this bound is asymptotically tight, with base equal to the golden ratio.


    \bigskip\noindent
    \textbf{AMS-MSC 2020:} 05C75, 05C40, 68R10, 05C85.
\end{abstract}

    \small
    \tableofcontents

\begin{figure}[h]
	\centering
	\includegraphics[width=\linewidth]{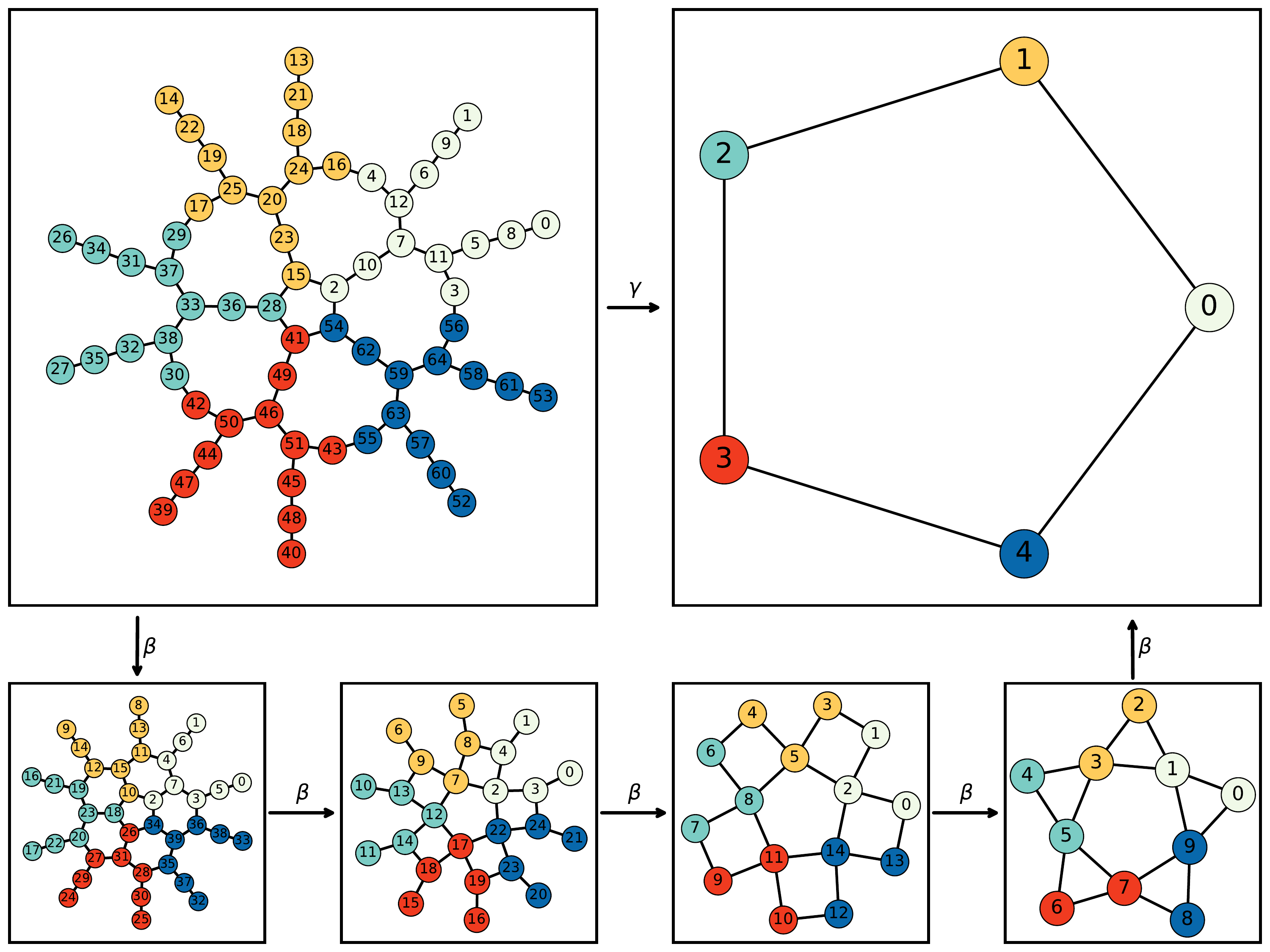}
	\caption{Example of the application of $\gamma$- and $\beta$-contraction when applied to a five-cluster graph. $\beta$-contraction iterativelly converges to the $\gamma$ contraction of the original graph in five steps (the example is carefully crafted to maximise the number of iterations for depiction purposes).}
	\label{fig:gamma-beta-approaches}
\end{figure}

\newpage

\section{Introduction}

Graphs play a central role in computer science and discrete mathematics, offering a way to synthesise relevant information from a wide variety of real-world scenarios in powerful abstract combinatorial structures~\cite{Diestel_25}.
A standard strategy for simplifying complex graphs consists in reducing their size by grouping vertices that share common features and replacing each group with a representative element~\cite{Ezugwu_Ikotun_Oyelade_etal_22}. 
The resulting reduced graph is hence smaller and more manageable than the original one, while preserving selected structural properties.

A natural and widely applicable way to define such groups relies on a categorical classification of vertices, \ie a mapping from the vertex set to a finite set of categories. 
In graph-theoretic terms, this corresponds to a (generally non-proper) vertex colouring. 
Such colourings may be obtained algorithmically, for instance via clustering techniques such as hierarchical clustering or highly connected subgraph methods~\cite{Xu_WunschII_05}. 
In many settings, however, categorical information is intrinsic to the graph and provided \emph{a priori}. 
Examples include semantic annotations in knowledge graphs or ontologies~\cite{Hogan_Blomqvist_Cochez_etal_21}, typological classifications injected by expert knowledge~\cite{Onofri_Corbetta_22}, or externally defined roles assigned to vertices in annotated networks. 
In these contexts, colours encode semantic information that should be preserved under any graph-reduction procedure.

From a mathematical perspective, reducing a graph according to a given classification naturally leads to the notion of a \emph{quotient graph}~\cite{Godsil_Royle_01}. 
Given a partition of the vertex set, a quotient graph identifies each block of the partition with a single vertex and preserves adjacency relations between blocks whenever at least one edge exists between their elements. 
Quotient constructions are classical and provide an abstract framework for graph reduction. 
However, when partitions are induced by categorical data, an additional structural requirement becomes essential: vertices should be identified only when they belong to the same connected region of equal category. 
Ignoring connectivity may lead to quotient graphs that obscure the original topology and collapse structurally unrelated regions.

A concrete and flexible way to enforce this connectivity constraint is through graph contraction. 
By iteratively merging adjacent vertices that share the same categorical information, contraction produces a reduced graph whose vertices correspond to connected components of equal colour. 
Unlike centroid-based approaches commonly used in data clustering, contraction does not introduce artificial vertices, but preserves representatives derived directly from the original graph. 
Moreover, the contraction operation can be defined so as to retain additional information about each component, such as its size and order or detailed information regarding internal connectivity like, \eg, clique presence, degree distribution, or size of the minimal cut.

Graph contraction has been extensively studied as a problem in its own right, including computationally hard variants such as the NP-hard tree contraction problem~\cite{Asano_Hirata_83}. 
In the present work, however, contraction is considered from a different viewpoint: as a systematic, quotient-like procedure for reducing vertex-coloured graphs while preserving categorical connectivity. 
To the best of our knowledge, a rigorous mathematical formalisation of this process, explicitly accounting for both colouring and connectivity, is lacking in the literature.
We refer to this formalised construction as \emph{$\gamma$-contraction}, and denote by $G/\gamma$ the quotient-like graph obtained from a graph $G$ equipped with a colouring $\gamma$.

The notion of $\gamma$-contraction was originally introduced in~\cite{Lombardi_Onofri_22,Lombardi_Onofri_22A}, together with an efficient iterative algorithm. 
However, a precise mathematical definition of the underlying construction and of its algorithmic assumptions was not provided. 
The main objective of the present work is to fill this gap by developing a rigorous formal framework for colour-based graph contraction. 
Building on this framework, we introduce the notion of \emph{$\beta$-contraction}, a weaker greedy formulation that captures the algorithmic core of $\gamma$-contraction. 
This reformulation enables a detailed theoretical analysis, including proofs of correctness and convergence (see Theorem~\ref{theo:D-forest} and Theorem~\ref{theo:beta-convergence}), while preserving the quotient-based interpretation that motivates the entire approach.
A graphical example of $\gamma$- and $\beta$-contraction approaches is depicted in Figure~\ref{fig:gamma-beta-approaches} for reference.

\subsection{Related Work}

Graph contraction and graph reduction techniques have been extensively studied as tools for simplifying graph structures while preserving selected combinatorial properties. 
Their use spans classical graph-theoretic problems as well as algorithmic settings in which contraction-based formulations play a central role. 
Representative examples include contraction-based shortest-path algorithms~\cite{Delling_Goldberg_Nowatzyk_etal_13,Guattery_Miller_92}, the evaluation of graph polynomials such as chromatic and DP-colouring polynomials~\cite{West_01,Mudrock_22}, and the computation of graph invariants, including the number of spanning trees~\cite{Gross_Yellen_Anderson_18}.

From a structural combinatorial perspective, contraction operations play a central role in defining canonical representations and invariants of graphs. In particular, contraction–deletion frameworks have been used to characterise graph polynomials and other invariants through recursive reduction rules~\cite{Bollobas_Pebody_Riordan_00}, while structural studies of connectivity-preserving contraction operations further illustrate how contraction interacts with intrinsic combinatorial properties~\cite{Ando_Egawa_Kriesell_13,Martinov_82}. More broadly, structural investigations of connectivity under contraction formalise contraction as a fundamental operation for comparing graph structure, culminating in the well-quasi-ordering of graphs under the minor relation~\cite{Robertson_Seymour_04}.
Related ideas also appear in the construction of canonical tree-decompositions, which provide unique structural representations capturing connectivity or block structure in graphs~\cite{Carmesin_Gollin_17,Carmesin_Hamann_Miraftab_22}.
While these approaches establish contraction as a foundational tool for structural analysis, they do not address contraction driven by externally prescribed categorical information, nor do they focus on preserving adjacency relations between colour classes.

From a mathematical standpoint, graph reduction is closely related to the notion of quotient graphs, in which vertices are identified according to an equivalence relation and collapsed into equivalence classes while preserving adjacency between classes. Quotient constructions provide an abstract framework for studying vertex identifications~\cite{Godsil_Royle_01}, and related formalisation arises in the theory of graph homomorphisms, which connect structure, symmetry, and component behaviour under identification maps~\cite{hell2004graphs}.
However, standard quotient graphs do not impose structural constraints on equivalence classes, such as connectivity, and are typically studied independently of algorithmic constructions. 
As a consequence, they do not directly capture reduction processes induced by categorical vertex colourings when connectivity preservation is required.

A substantial body of work on graph reduction also originates from graph coarsening and multilevel methods, particularly in the context of parallel computing and large-scale optimisation~\cite{Valejo_Ferreira_Fabbri_etal_20}. 
In these approaches, graphs are iteratively reduced by aggregating vertices according to local heuristics, producing hierarchies of progressively smaller graphs. 
Early parallel contraction schemes were proposed to reduce problem size prior to data mapping~\cite{Ponnusamy_Mansour_Choudhary_etal_93}, and were later extended into multilevel frameworks combining contraction with techniques such as label propagation~\cite{Meyerhenke_Sanders_Schulz_14}.
While such methods share algorithmic similarities with contraction-based reduction, they are typically optimisation-driven and heuristic in nature, and the resulting reduced graphs are not intended as canonical representatives of the original structure.

Related notions of reduction and aggregation also appear in other areas, such as the study of stochastic processes and dynamical systems, where partitions of large state spaces are used to derive reduced models preserving behavioural properties~\cite{Geiger_26}, including reachability or long-term dynamics~\cite{Kemeny_Snell_76,Buchholz_94}. 
Despite conceptual similarities, these approaches rely, however, on transition dynamics or behavioural equivalence rather than on (potentially externally prescribed) fixed categorical information, and therefore differ from the scope of our colour-based graph contraction.

A complementary perspective is offered by approaches that analyse large networks via spanning or sampled subgraphs, often used to infer hidden structural properties~\cite{Ahmed_Neville_Kompella_13,DallAsta_Alvarez-Hamelin_Barrat_etal_06,Bernaschi_Celestini_Guarino_etal_17}.
In such studies, structural properties are inferred from selected subgraphs rather than from an explicit reduction of the full graph, allowing statistical characterisation of features such as connectivity, centrality, or community structure. While effective for exploratory or measurement-driven analyses, these methods do not provide a canonical compressed representation of the network, and the relationships between distinct regions of the original graph can only be partially recovered. In contrast, our colour-based contraction produces a reduced graph that preserves adjacency amongst distinct colour classes, offering a deterministic summarisation of the full network topology.

Only a limited number of works explicitly address graph contraction driven by semantic or categorical vertex annotations. 
Examples include contraction-based analyses of urban and architectural graphs~\cite{DAutilia_Spada_18,DAutilia_Marrone_Palma_25,Onofri_Corbetta_22}, social and web networks~\cite{Lombardi_Onofri_22}, and other annotated graph structures. 
These contributions highlight the usefulness of colour-based contraction as a means of structural abstraction, but do not provide a unified mathematical formalisation of the underlying contraction process.

\subsection{Paper contributions}

The main contributions of this work can be summarised as follows:
\begin{itemize}
    \item We introduce a rigorous mathematical formalisation of colour-based graph contraction, termed \emph{$\gamma$-contraction}, modelling graph reduction as a quotient-like operation induced by a fixed vertex colouring and constrained by connectivity.
    To the best of our knowledge, this is the first formal treatment of colour-based contraction that explicitly enforces connectivity of colour classes.

    \item We develop a coherent theoretical framework and notation for colour-induced contraction, clarifying the relationship between $\gamma$-contraction, classical graph contraction, connectivity, and quotient graph constructions.

    \item We define a weaker, local formulation of colour-based contraction, called \emph{$\beta$-contraction}, which replaces global graph traversals with vertex-local conditions.
    This formulation captures the core mechanism underlying $\gamma$-contraction and provides a constructive characterisation of the contraction process (Theorem~\ref{theo:D-forest}).

    \item Building on the theory of $\beta$-contraction, we derive a general algorithmic strategy for computing $\gamma$-contractions and introduce a concrete contraction algorithm with provable correctness and convergence guarantees.
    Although presented in a serial form, the algorithm follows design principles that are compatible with parallel graph processing and avoid global synchronisation steps.

    \item We establish formal guarantees on the convergence of $\beta$-contraction, proving that the contraction of a connected colour component completes in at most $\lfloor \log_\varphi(n) \rfloor$ iterations, where $\varphi$ denotes the golden ratio (Theorem~\ref{theo:beta-convergence}).
\end{itemize}

\subsection{Paper organisation}

The rest of the paper is organised as follows.
Section~\ref{sec:prerequisites} recalls basic notions and notation from graph theory, with particular emphasis on classical graph contraction (Section~\ref{ssec:graph-contraction}).
Section~\ref{sec:colour-contraction} introduces the formal definition of colour-based graph contraction and establishes the theoretical foundations of $\gamma$-contraction.

Section~\ref{sec:beta-contraction} is devoted to the study of $\beta$-contraction.
After providing a structural overview of the construction (Section~\ref{ssec:beta-overview}), we analyse its convergence properties and prove logarithmic convergence bounds for connected colour components (Section~\ref{subsec:beta-convergence}).

Section~\ref{sec:algorithmic-framework} presents the algorithmic realisation of $\beta$-contraction.
We first introduce the data structures used to represent graphs and contraction mappings (Section~\ref{subsec:data-structures}), then provide a high-level overview of the algorithmic framework (Section~\ref{subsec:algo-overview}).
The detailed construction of the contraction mapping and its application to the graph are discussed in Sections~\ref{ssec:beta-construction} and~\ref{ssec:beta-application}, respectively, followed by an analysis of the computational complexity (Section~\ref{ssec:complexity-analysis}).

Finally, Section~\ref{sec:conclusions} summarises the main results and outlines directions for future work.
The formal proof of the convergence theorem and the im\-ple\-men\-ta\-tion-level details are deferred to the appendices.

\subsection{General notation}

Throughout the paper, capital Latin letters denote sets, lower-case Latin letters denote individual objects, Greek letters denote mappings, Gothic letters denote operators, and typewriter font is used for algorithm-related variables.
Notation is used consistently; for instance, $u, v, w$ always denote vertices.
Asymptotic bounds are expressed using the standard big-$\OOO$ and big-$\Omega$ notation: we write $f(x) \in \OOO(g(x))$ [resp.\ $f(x) \in \Omega(g(x))$] if there exist constants $c > 0$ and $x_0 < \infty$ such that $f(x) \le c\,g(x)$ [resp.\ $f(x) \ge c\,g(x)$] for all $x \ge x_0$.
Disjoint union is denoted with $\sqcup$, namely, given two sets $U, V$, we define $U \sqcup V = (U \cup V) \setminus (U \cap V)$.


\section{Prerequisites on graph}\label{sec:prerequisites}

\subsection{Basic definitions}

Given a set of vertices (or nodes) $V$, we recall from the basics in graph theory the definition of a \emph{graph} as a couple $G = (V, E)$, where $E \subset V \times V$ represents a set of edges connecting the nodes two by two.
In other words, $E$ can be seen as a relation $\rho_G$ (or $\rho_E$) on $V$, where we say that two vertices $u, v \in V$ are adjacent if $u \rho_G v$ or/and $v \rho_G u$, \ie $(u, v) \in E$ or/and $(v, u) \in E$.
In particular, we refer to the graph as \emph{undirected} if $\rho_G$ is symmetric, \ie $(u, v) \in E \iff (v, u) \in E$, and we refer to it as \emph{simple} if the relation is co-reflexive, \ie $(u, u) \not\in E, \forall u \in V$. In simple undirected graphs, we denote edges as 2-sets $\{u, v\}$ instead of ordered pairs $(u, v)$. We define the \emph{order} of a graph $G$ as the size of its vertex set and the \emph{size} of $G$ as the size of its edge set; we denote them as $n_G = |G| = |V_G|$ and $m_G = \|G\| = |E_G|$ respectively.

We define the \emph{neighbourhood} $N(u)$ of a vertex $u \in V$ as the set of vertices adjacent to $u$, \ie $N_G(u) = \{v \in V \mid u \rho_G v \lor v \rho_G u\}$.
Analogously, for the sake of convenience, we also extend the notion of neighbourhood to any set of vertices $U \subset V$ as the set of vertices adjacent to at least one vertex $u \in U$, \ie $N_G(U) = \big(\bigcup_{u \in U} N_G(u)\big)\backslash U$.
The size of $N(v)$ is referred to as \emph{degree} of $v$ and it is denoted by $\degree_G(v)$.
If a directed graph is considered, we identify with $N^\scin(v)$, $N^\scout(v)$, $\degree^\scin$, and $\degree^\scout$ the set of vertices reachable with edges ending up or starting from $v$ and the corresponding sizes.

In what follows, most of our results are presented in terms of simple undirected graphs; however, they naturally extend also to non-simple and/or directed graphs, as it can be seen in \cite{Caprolu_Di-Pietro_Lombardi_etal_24}.
For this reason, we refer to a simple undirected graph simply as a graph, if not stated differently, and, for the sake of readability, we denote $\rho_G$ by $\sim_G$ to recall its symmetric property.
Conversely, we refer to a directed graph as a di-graph, and we adopt $\rightsquigarrow_G$ as the edge relation.
For the sake of readability, we also drop the subscripts when they can be inferred from the context.

We recall that $H = (S, F)$ is a \emph{subgraph} of $G = (V, E)$, and we write $H \leq G$, if $S \subset V$ and $F \subset E$. In particular, we say that $H$ is the subgraph of $G$ \emph{induced} by $S$ if it holds that $u \sim_G v \Rightarrow u \sim_H v, \forall u, v \in S$; in this case, we denote it as as $H = \langle S \rangle_G$ and we say that $S$ \emph{spans} (or induces) $H$ over $G$. Finally, we say that $G'$ is a \emph{spanning graph} if $G' \leq G$ and $\langle G'\rangle_G = G$, \ie $V' = V$.
Graphs spannings conveniently defines the removal of a set of vertices $U \subset V$ from $G$ as $G-U = \langle V\backslash U\rangle_G$.
Conversely, given a set of edges $F \subset V\times V$, we can naturally define the addition [resp.\ removal] of said set to [resp.\ from] $G$ as $G+F = (V, E\cup F)$ [resp.\ ($G-F = (V, E \backslash F)$].
Finally, given a second graph $G_1 = (V_1, E_1)$, we define the union graph as $G \cup G_1 = (V \cup V_1, E \cup E_1)$.

A (simple) \emph{path} $\Gamma_{u, v}$ between two vertices $u, v$ is a sequence of distinct vertices $w_0, w_1, \dots w_\ell$, where $w_0 = u$, $w_\ell = v$ and, for $0 \leq i < \ell$, we have $w_{i+1} \in N_G(w_{i})$.
Two nodes are said to be \emph{connected} if there exists a path between them; otherwise, they are \emph{disconnected}.
Analogously, a graph is referred to as \emph{connected} if each pair of nodes is mutually connected.
A connected subgraph which is maximal is said to be a \emph{connected component} while the set of all the connected components of a graph is unique and it is referred to as \emph{partition}.

If for any couple of vertices there exists at most one path connecting them, then the graph is said a \emph{forest}; if there exists exactly one path connecting each couple of vertices, then the graph is said to be a \emph{tree}.
Given a tree $T = (V, E)$, it is always possible, and sometimes convenient, to select a given node $r \in V$ as a special node; such node is then called the \emph{root} of the tree.
A tree with a fixed root is often called \emph{rooted tree} -- denoted by $T_r$.

Vertices or/and edges of a graph can be, in general, equipped with any sort of data, hence forming a weighted graph.
In the rest of the paper, we mainly restrict these kinds of data to categorical information -- data sampled from a numerable finite set -- over vertices, often referred to as (vertex) colours.
More formally, given a set of colours $C$, $|C| < \infty$, a (vertex-) \emph{colouring} is a map $\gamma: V \to C$ that associates every vertex with a colour.
A coloured graph is hence formed by a triple $G = (V, E, \gamma)$.
It is important to notice that, in the classical literature of graph theory, many efforts were devoted to studying proper graph colourings -- often abbreviated simply as colourings -- where $\gamma(u) \ne \gamma(v)$, $\forall \{u, v\} \in E$.
This is not the context of the present work, since we are mainly interested in analysing contiguous \emph{patches} of colour, or coloured components, as we will see in the next section.

\subsection{Graph contraction}\label{ssec:graph-contraction}

A common strategy for analysing large graphs consists of replacing the original graph with a smaller structure that preserves essential connectivity information.
One of the most classical operations serving this purpose is \emph{graph contraction}.

Given a graph $G = (V,E)$ and two adjacent vertices $u,v \in V$, we define the \emph{contraction} of $u$ and $v$ --and we write $G/\{u,v\}$-- as the graph obtained by identifying $u$ and $v$ into a single new vertex $w$ and preserving all adjacencies to the remaining vertices.
Formally, the contracted graph $G'=(V',E')$ is defined by
\begin{align}
        &V' = (V \backslash \{u, v\}) \cup \{w\}
        \\
        &E' = \{\{x, y\} \mid x, y \in V', \: x \sim_E y\}\; \cup\; \overbrace{\{\{x,w\} \mid x\in N_G(\{u, v\})\}}^{E_w}\ ,
\end{align}
or, analogously,
\begin{equation}\label{gt-eq:edge-contraction}
	G/\{u, v\} = \Big(\big(G - \{u, v\}\big) \cup \big(\{w\}, \emptyset\big) \Big) + E_w\ .
\end{equation}
Figure~\ref{gt-fig:regular-contraction} illustrates this operation.

It is worth noticing that $w$ is somehow derived by $u$, $v$, and the edge $\{u, v\}$, as well as edges $\{x, w\}$ are derived by [the combination of] $\{x, u\}$ or [and] $\{x, v\}$, $\forall x \in N_G(\{u, v\})$. As an example, in the context of weighted graphs, vertices and edges might inherit a combination of the original weights, and the weight of $\{u, v\} \in E$ might \eg be transferred to the vertex $w$.

\begin{figure}[tp]
    \centering
    \begin{subfigure}{.49\linewidth}
        \centering
        \begin{tikzpicture}[every node/.style={draw,fill,circle}]
    \node[] (u) at (0, 0) {};
    \node[] (v) at (0, 1) {};
    \node[fit=(u)(v), ellipse, inner xsep=+2pt, inner ysep=0pt, opacity=.2] {};
    
    \node[label=below:$u$, fill=orangeM] at (0, 0) {};
    \node[label=above right:$v$, fill=orangeM] at (0, 1) {};
    \node[] (0) at (-0.5, -0.8) {};
    \node[] (1) at (-0.7, +0.3) {};
    \node[] (2) at (-1.2, +1.1) {};
    \node[] (3) at (-0.1, +1.8) {};
    \node[] (4) at (+1.5, +1.3) {};
    \node[] (5) at (+0.7, +0.6) {};
    \node[] (6) at (+1.8, +0.3) {};
    \node[] (7) at (+0.8, -0.3) {};
    \node[] (8) at (-1.4, -0.5) {};
    \draw[-, thick] (u) -- (v);
    \draw[-, thick]
        (u) -- (0) -- (8)
        (u) -- (1) -- (8)
        (1) -- (2)
        (v) -- (2) -- (3)
        (v) -- (5) -- (4) -- (3)
        (u) -- (7) -- (6) -- (4)
        (5) -- (7)
    ;
\end{tikzpicture}
        \caption{}
        \label{gt-fig:regular-contraction-a}
    \end{subfigure}
    \begin{subfigure}{.49\linewidth}
        \centering
        \begin{tikzpicture}[every node/.style={draw,fill,circle}]
    \node[label=above:$w$, fill=greenM] (w) at (0, 0.5) {};
    \node[] (0) at (-0.5, -0.8) {};
    \node[] (1) at (-0.7, +0.3) {};
    \node[] (2) at (-1.2, +1.1) {};
    \node[] (3) at (-0.1, +1.8) {};
    \node[] (4) at (+1.5, +1.3) {};
    \node[] (5) at (+0.7, +0.6) {};
    \node[] (6) at (+1.8, +0.3) {};
    \node[] (7) at (+0.8, -0.3) {};
    \node[] (8) at (-1.4, -0.5) {};
    \draw[-, thick]
        (w) -- (0) -- (8)
        (w) -- (1) -- (8)
        (1) -- (2)
        (w) -- (2) -- (3)
        (w) -- (5) -- (4) -- (3)
        (w) -- (7) -- (6) -- (4)
        (5) -- (7)
    ;
\end{tikzpicture}
        \caption{}
        \label{gt-fig:regular-contraction-b}
    \end{subfigure}
    \caption[Sample of graph contraction]{
        \subref{gt-fig:regular-contraction-a} a graph $G$ and \subref{gt-fig:regular-contraction-b} the corresponding contraction $G' = G/\{u, v\}$. Contracted vertices $u, v \in V_G$ are highlighted in orange and the resulting blended vertex $w \in V_{G'}$ is highlighted in green. 
    }
    \label{gt-fig:regular-contraction}
\end{figure}

From an operative perspective, we can say that the contraction is an operator $\mmmm(G, u, v) \equiv\mmmm_G(u, v)$ that maps $G$ to the contracted graph $G' = G/\{u, v\}$, obtained according \eqref{gt-eq:edge-contraction}; the blended vertex $w\not\in V$ is then obtained from $u$ and $v$ under some blending function that might involve $\{u, v\} \in E$ as well, and the set $E_w$ of blended edges is created so that $w$ is connected to $G-\{u, v\}$ preserving the original adjacencies: here edges from $N(u) \sqcup N(v)$ are simply rewired to $w$ while novel blended edges combining $\{u, z\}$ and $\{v, z\}$ are built whenever $z \in N(u) \cap N(v)$.
For convenience, in what follows we drop the subscript $G$ from $\mmmm_G$ when it can be inferred from the context.

In particular, $\mmmm$ satisfies commutative [$\mmmm(u, v) = \mmmm(v, u)$] and associative [$\mmmm(u, \mmmm(v, w)) = \mmmm(\mmmm(u, v), w)$] properties if the same properties hold for the above-mentioned blending functions. Here, we are assuming $w \in N_G(\{u, v\})$ and we are denoting the iterated application of contraction with the shortened form $\mmmm(\mmmm(u, v), w)$ instead of the more thorough yet tedious notation ``$(\mmmm_{\bar G}(\bar x, w)$ where $\bar G = G/\{u,v\}$ and $\bar x$ being the resulting blended node from $\mmmm_G(u, v)$''.
In what follows, we assume the two properties hold true (see Remark\ref{gt-fn:gc-associative-commutative}).
It follows that $\mmmm$ can be seen as a \emph{variadic} operator, \ie it can be applied to any set of vertices $U = \{u_0, \dots, u_k\} \subset V$ that spans a connected subgraph $\langle U \rangle_G$ --and we denote this with $\mmmm_G(U)$.
In other words, we have that $\mmmm_G(U) = \mmmm_G(u_0, \dots, u_k) =  \mmmm(\dots(\mmmm(u_0, u_1)\dots), u_k)$ for some ordering of the vertices like, \eg, the ordering induced by a spanning tree of $\langle U \rangle_G$.

\begin{remark}[On the properties of $\mmmm$]\label{gt-fn:gc-associative-commutative}
	Here, do notice that the assumption over associative and commutative properties for $\mmmm$ naturally hold true for a simple unweighted graph since colour is preserved but, more in general, we can distinguish the contraction of $u$ in $v$ from the contraction of $v$ in $u$, depending on how additional data are generated for $w$ and its corresponding edges.
	However, such an outcome is out of the scope of this work, since we will only consider additive or invariant blending functions over weighted graphs, \ie where $\omega_V(w) = \omega_V(u) + \omega_V(v)$, with $\omega_V:V\to \RR$ a vertex weight function, or where $\gamma_V(w) = \gamma_V(u) = \gamma_V(v)$, with similar behaviour on edges.
\end{remark}

More generally, provided the blending function is associative and commutative, contraction can be extended from pairs of vertices to arbitrary connected subsets of vertices.
Let $U \subset V$ be a set of vertices that induces a connected subgraph of $G$.
The \emph{contraction of $U$}, denoted by $G/U$, is the graph obtained by identifying all vertices in $U$ into a single blended vertex $w$, with adjacency defined under a set of (potentially) blended edges $E_U$, that is
\[
G/U = \big( (G - U) \cup (\{w\}, \emptyset) \big) + \overbrace{\{\{x,w\} \mid x \in N_G(U)\}}^{E_U} \,.
\]

\begin{figure}[tp]
    \centering
    \begin{subfigure}{.49\linewidth}
        \centering
                \begin{tikzpicture}[vert/.style={draw,fill,circle}]
            \node[vert] (u) at (0, 0) {};
            \node[vert] (v) at (0, 1) {};
            \node[vert] (z) at (-0.7, +0.3) {};
            \node[vert, fit=(u)(v)(z), ellipse, inner xsep=+0pt, inner ysep=0pt, opacity=.2, rotate=-20, xshift=.1cm] {};
            \node[] at (-.3,.5) {$U$};
            
            \node[vert, fill=orangeM] at (0, 0) {};
            \node[vert, fill=orangeM] at (0, 1) {};
            \node[vert, fill=orangeM] at (-0.7, +0.3) {};
            \node[vert] (0) at (-0.5, -0.8) {};
            \node[vert] (2) at (-1.2, +1.4) {};
            \node[vert] (3) at (-0.1, +1.8) {};
            \node[vert] (4) at (+1.5, +1.3) {};
            \node[vert] (5) at (+0.7, +0.6) {};
            \node[vert] (6) at (+1.8, +0.3) {};
            \node[vert] (7) at (+0.8, -0.3) {};
            \node[vert] (8) at (-1.4, -0.5) {};
            \draw[-, thick] (u) -- (v);
            \draw[-, thick]
                (u) -- (0) -- (8)
                (u) -- (z) -- (8)
                (z) -- (2)
                (v) -- (2) -- (3)
                (v) -- (5) -- (4) -- (3)
                (u) -- (7) -- (6) -- (4)
                (5) -- (7)
            ;
        \end{tikzpicture}
        \caption{}
        \label{gt-fig:variadic-contraction-a}
    \end{subfigure}
    \begin{subfigure}{.49\linewidth}
        \centering
        \begin{tikzpicture}[every node/.style={draw,fill,circle}]
    \node[label=above:$w$, fill=greenM] (w) at (-.23, 0.43) {};
    \node[] (0) at (-0.5, -0.8) {};
    \node[] (2) at (-1.2, +1.4) {};
    \node[] (3) at (-0.1, +1.8) {};
    \node[] (4) at (+1.5, +1.3) {};
    \node[] (5) at (+0.7, +0.6) {};
    \node[] (6) at (+1.8, +0.3) {};
    \node[] (7) at (+0.8, -0.3) {};
    \node[] (8) at (-1.4, -0.5) {};
    \draw[-, thick]
        (w) -- (0) -- (8)
        (w) -- (8)
        (w) -- (2) -- (3)
        (w) -- (5) -- (4) -- (3)
        (w) -- (7) -- (6) -- (4)
        (5) -- (7)
    ;
\end{tikzpicture}
        \caption{}
        \label{gt-fig:variadic-contraction-b}
    \end{subfigure}
    \caption[Sample of graph contraction (variadic form)]{
        \subref{gt-fig:variadic-contraction-a} a graph $G$ and \subref{gt-fig:variadic-contraction-b} the corresponding contraction $G' = G/U$. The set of contracted vertices $U \subseteq V_G$ is highlighted in orange and the resulting blended vertex $w \in V_{G'}$ is highlighted in green. 
    }
    \label{gt-fig:variadic-contraction}
\end{figure}

We refer to $G/U$ as a \emph{contraction graph} of $G$ and write $G/U \preceq G$.
If $|U| \ge 2$, the contraction is said to be \emph{proper} and we write $G/U \prec G$.
Figure~\ref{gt-fig:variadic-contraction} provides an example of contraction of a vertex set (cf.\ Figure~\ref{gt-fig:regular-contraction}).

Commutative and associative properties of the blending functions also allow extending contraction to multiple disjoint sets of connected vertices, \ie given a graph $G = (V, E)$ and two sets of nodes $U, W \subset V$, with $U \cap W = \emptyset$, then $(G/U)/W$ and $(G/W)/U$ produce the same graph. Consequently, it makes sense to drop the parentheses, hence simply writing $G/U/W$, and to consider the extension to any number of independent sets of connected nodes.
However, we differ $G/U/W$ from $G/(U\cup W)$ by the number of blended vertices we are introducing in the contracted graph, as $G/U/W$ produces one blended node for $U$ and one for $W$ while $G/(U\cup W)$ produces a single node and requires $\exists u \in U, v \in V$ such that $\{u, v\} \in E$.
\begin{remark}[On the set-disjoint requirement of progressive contractions]
	Here, do notice that the disjoint requirement of $U \cap W = \emptyset$ while defining $G/U/W$ allows for consistency in the notation, as otherwise it would not be possible to deceive the number of created blended vertices.
\end{remark}

From a theoretical perspective, contraction operations can be related to the notion of \emph{quotient graphs}. Given an equivalence relation $\sigma$ on the vertex set $V$, the quotient graph $G/{\sigma}$ is defined as the graph whose vertices are the equivalence classes $V_\sigma$, with an edge between two classes whenever at least one edge exists in $G$ between their representatives.
Quotient graphs provide a compact mathematical formalisation of vertex identification and are widely used in graph theory to reason about symmetry and structural reduction.
However, in their most general form, quotient constructions do not impose any connectivity requirement on equivalence classes, nor do they specify how equivalence relations should be generated. In particular, if equivalence classes are not connected subgraphs of $G$, the resulting quotient may collapse vertices that are structurally distant, potentially obscuring relevant topological information.

The contraction operations introduced above can hence be interpreted as quotient constructions induced by equivalence relations whose classes are required to span connected subgraphs.
In this sense, variadic contraction $G/U$ corresponds to the quotient of $G$ with respect to the equivalence relation that identifies precisely the vertices in $U$ and leaves all other vertices distinct.
This connectivity-aware interpretation of quotienting will play a central role in the definition of colour-based contraction developed in the following sections.


\section{Coloured-based graph contraction}\label{sec:colour-contraction}

In this section, we introduce a contraction operation for vertex-coloured graphs that is sensitive not only to vertex labels, but also to the underlying graph topology: the $\gamma$-contraction.
The goal is to collapse vertices sharing the same colour while preserving adjacency information between distinct colour regions, without introducing artificial representatives or optimisation criteria.

At an intuitive level, $\gamma$-contraction operates by collapsing each maximal connected region of vertices sharing the same colour into a single representative node.
The key point is that colour alone is not sufficient to induce contraction: vertices are merged only if they are connected through paths entirely composed of vertices of the same colour.
As a consequence, vertices with identical categorical labels but belonging to disconnected regions of the graph are kept distinct in the contracted representation.

This connectivity-aware identification distinguishes $\gamma$-contraction from both standard clustering techniques and purely quotient-based constructions.
In particular, no optimisation criterion is involved, no artificial representatives (such as centroids or medoids) are introduced, and the contraction outcome depends solely on the interplay between graph topology and categorical information.
The resulting contracted graph can therefore be seen as a semantic projection of the original structure, preserving inter-class adjacency while compressing intra-class connectivity.

\subsection{Colour-preserving contraction}

Let $G = (V, E, \gamma)$ be a graph equipped with a vertex-colouring function $\gamma : V \to C$, where $C$ is a finite set of colours.
Unlike quotient constructions based solely on equivalence relations, the contraction mechanism proposed here identifies vertices only when they are connected through paths entirely composed of vertices of the same colour.
As a consequence, vertices with identical colours but belonging to disconnected regions of the graph remain distinct after contraction.

This connectivity-aware approach yields a canonical reduced graph whose vertices correspond to maximal monochromatic connected subgraphs of $G$.
We formalise this construction by first extending the classical notion of graph contraction to the coloured setting, and then introducing the notion of $\gamma$-contraction as a variadic contraction over colour components.

\begin{definition}[Colour-preserving contraction]\label{def:colourPreserving}
    Let $G=(V, E, \gamma)$ be a vertex-coloured graph and let $u, v \in V$ be two adjacent vertices sharing the same colour, \ie $u \sim v$ and $\gamma(u) = \gamma(v)$.
    We define the \emph{colour-preserving contraction} --and we denote it as $G/_\gamma\{u, v\}$-- as the graph $G' = G/\{u, v\}$ equipped with the colouring function
    \begin{equation}\label{eg:colour-preserve}
        \begin{matrix}
            \gamma' & : & V' & \to & C \\
            & & z & \mapsto & \gamma'(z)
        \end{matrix}
        \ , \qquad \text{with} \qquad
        \gamma'(z) =
        \begin{cases}
            \gamma(z) & \mbox{if } z \ne w\\
            \gamma(u) & \mbox{if } z = w
        \end{cases}
    \end{equation}
    where $w$ is the novel vertex in $G'$ that blends $u$ and $v$.
    
    We define the corresponding operator as $\mmmm_G^\gamma(u, v)$ analogously to $\mmmm_G(u, v)$, mapping $G$ equipped with $\gamma$ to $G/\{u, v\}$ equipped with $\gamma'$.
\end{definition}

It is then straightforward to consider the following.
\begin{proposition}[Properties of colour-preserving contraction]\label{prop:commasscont}
    Colour-preserving contraction $\mmmm_G^\gamma$ is commutative and associative.
\end{proposition}
\begin{proof}
    We proceed proving commutative property. Associative property follows analogously.
    
    Let us denote with $H$ and $H'$ the graphs obtained under $\mmmm_G^\gamma(u, v)$ and $\mmmm_G^\gamma(v, u)$ respectively.
    Let us also denote with and with $w$ and $w'$ the corresponding blended vertices, incident on $E_w$ and $E_{w'}$, equipped with colour $c_w$ and $c_{w'}$ resp.
    In particular, we already know by the definition of $\mmmm_G$ that $H=H'$, $w=w'$, and $E_w = E_{w'}$.
    From \eqref{eg:colour-preserve}, it follows $c_w = \gamma(u)$ and $c_{w'} = \gamma(v)$.
    Since colour-preserving contraction is solely defined on $u, v$ such that $\gamma(u) = \gamma(v)$ we can conclude that $c_w=c_{w'}$.
\end{proof}

In general, we refer to $\mmmm_G^\gamma$ simply as $\mmmm$ when it is clear from the context that we are considering a coloured graph.

We can now provide a first intuitive definition of $\gamma$-contraction.

\begin{definition}[$\gamma$-contraction]\label{def:colourContraction_v1}
    Given a vertex-coloured graph $G= (V, E, \gamma)$, we define the \emph{$\gamma$-contraction} of $G$, more formally the \emph{colour contraction} of $G$ on the colour set $C$ induced by the colouring $\gamma$, as the procedure that applies the (commutative, associative) colour-preserving graph contraction to all the couples of adjacent vertices that share the same colour, \ie that evaluates progressively $G/_\gamma\{u, v\}, \forall u, v \in V$ such that $u\sim v$ and $\gamma(u) = \gamma(v)$.
    We denote such an operation as $G/\gamma$.
\end{definition}

\subsection{Colour clusters and colour components}

Before proceeding, it is worthwhile for ease of notation to provide the following set of definitions on coloured graphs.
\begin{definition}[Colour neighbourhood]\label{def:colourNeighbourhooh}
    Let $G = (V, E, \gamma)$ be a vertex-coloured graph.
    We define the \emph{colour neighbourhood} of $v\in V$ -- and we write $N_\gamma(v)$ -- as the set of adjacent vertices of $v$ that share the same colour as it
    \begin{equation}
        N_\gamma(v) \equiv N_{G, \gamma}(v) = \{u \in N_G(v) \mid \gamma(u) = \gamma(v)\} \ .
    \end{equation}
    Analogously to the regular neighbourhood, we define the colour neighbourhood of a set of vertices $U \subseteq V$ as
    \begin{equation}\label{eq:colourNeighbourhood-U}
        N_\gamma(U) = \bigcup_{u \in U} N_\gamma(u) \backslash U\ .
    \end{equation}

    We refer to the size of the colour neighbourhood as the vertex/vertex set \emph{colour degree}, \ie
    \begin{equation}
        \degree_\gamma(u) = |N_\gamma(v)|\ , \qquad \degree_\gamma(U) = |N_\gamma(U)|\ .
    \end{equation}
\end{definition}

\begin{remark}[On colour neighbourhoods]
    In the following, we assume that whenever referring to $N_\gamma(U)$, we have $U$ is monochromatic, \ie $\gamma(u) = \gamma(v), \forall u, v \in U$. In fact, dropping such a condition would result in having multiple colours amongst the vertices in $N_\gamma(U)$.
\end{remark}

\begin{definition}[Colour cluster and colour component]\label{def:colourClusterComponent}
    Let $G = (V, E, \gamma)$ be a vertex-coloured graph.
    We define a \emph{colour cluster} as a set of vertices $S \subseteq V$ such that $\forall u, v \in S$ we have $\gamma(u) = \gamma(v)$ and there exists $\Gamma_{u, v} = (w_0, \dots, w_k)$ entirely contained in $S$, \ie $w_i \in S$, for $0 \leq i \leq k$.
    We denote with $\gamma(S)$ the colour of the vertices contained in $S$.
    
    If a colour cluster $S$ is \emph{maximal} (\ie $\forall S' \subset V$ colour cluster, $S \subseteq S'$ implies $S=S'$) we say that $S$ is a \emph{colour component}.
\end{definition}

\begin{remark}[On the connectivity of colour clusters]
    Requiring the existence between $u$ and $v$ of $\Gamma_{u, v}$ entirely contained in $S$ is equivalent to requiring that $\langle S\rangle_G$ is connected.
    In Definition~\ref{def:colourClusterComponent}, we explicitly refer to the path $\Gamma$ joining each couple of vertices in $S$ as it is later useful in both the proof of Proposition~\ref{prop:colourComponent} and in the explanation of the main algorithm introduced in this work.
\end{remark}

A useful remark on colour components, also highlighting the origin of the name that is rooted with their similarity with connected components, is provided in the following.
\begin{remark}[Colour components are disjoint]\label{rm:colourComponentsDisjoint}
    It directly follows from Definition~\ref{def:colourClusterComponent} that, given two colour components $S_1, S_2$ of $G$, then if $v \in S_1$ and $v \in S_2$ we have $S_1 = S_2$.
    In other words, $S_1 \cap S_2$ is either empty or equal to $S_1$ (and $S_2$).
\end{remark}

One more useful definition follows.
\begin{definition}[Colour (sub-)partition]
    Let $G = (V, E, \gamma)$ be a vertex-coloured graph.
    We define a \emph{colour sub-partition} of $G$ as a set of colour clusters $\SSS$ that forms a partition on the vertex set $V$.
    We further define a \emph{colour partition} $\SSS^\star$ as a minimal (sized) colour sub-partition.
\end{definition}
It is pretty straightforward to claim the following.
\begin{proposition}[The colour partition is unique]
    Let $G = (V, E, \gamma)$ be a vertex-coloured graph.
    The colour partition $\SSS^\star$ of $G$ -- and we denote it as $\CCC_\gamma(G)$ -- is unique and it is made of all the colour components of $G$.
\end{proposition}
\begin{proof}
    The proof follows from \remarkname~\ref{rm:colourComponentsDisjoint}.
    Since each vertex is contained in a single colour component, then the set $\SSS^\star$ of all the colour components of $G$ is unique.
    Furthermore, $\forall S_i, S_j \in \SSS^\star, S_i \ne S_j$, it holds $S_i \cap S_j = \emptyset$ and, since $S^\star$ collects all the colour components, then $\forall v \in V, \exists!S \in \SSS^\star$ such that $v \in S$, \ie $\SSS^\star$ is a partition of $V$.
    Minimality directly follows from the fact that colour components are maximal. 
\end{proof}

In order to define colour partitions in a constructive fashion, it is useful to characterise the colour components by means of the following.
\begin{proposition}[Colour component characterisation]\label{prop:colourComponent}
    Let $G = (V, E, \gamma)$ be a vertex-coloured graph and let $S\subseteq V$ be a colour cluster of $G$.
    We have that $S$ is a colour component $\iff$ $N_\gamma(S) = \emptyset$.
\end{proposition}
\begin{proof} We prove one implication at a time, both \adabsurdum.
    \begin{description}
        \item[($\Leftarrow$)] Let $S \subseteq V$ be a colour cluster such that $N_\gamma(S) = \emptyset$. If \adabsurdum $\exists S' \subseteq V$ such that $S'$ is a colour cluster and $S \subsetneq S'$, then $\hat S = S' \backslash S \ne \emptyset$.
        In particular, let us consider $v \in \hat S$ and $u \in S$ then $\exists \Gamma_{u, v} = (w_0, \dots, w_\ell)$, with $w_i \in S', \forall 0 \leq i \leq \ell$ and $w_0 = u$, $w_\ell = v$.
        Let us denote with $\hat i$ the first $i$ such that $w_i \not \in S$, \ie
        \[
            \hat i = \min \{i \in \{0, \dots, \ell\} \mid w_i \in \hat S\}\ .
        \]
        We have that $w_{\hat{i}-1}\in S$ (which is well-posed, in fact $\hat i > 0$ since $w_0 = u \in S$), hence $\{w_{\hat{i}-1}, w_{\hat{i}}\}\in E$ and $\gamma(w_{\hat{i}-1}) = \gamma(w_{\hat{i}})$ since $w_{\hat{i}-1}, w_{\hat{i}} \in S'$.
        It follows that $w_{\hat{i}} \in N_\gamma(w_{\hat{i}-1}) \subset N_\gamma(S) \ne \emptyset$ \absurd.

        \item[($\Rightarrow$)] Let $S \subset V$ be a colour component.
        If \adabsurdum $\exists u \in N_\gamma(S)$, let us prove that $S' = S\cup \{u\}$ is a colour cluster (such that $S \subsetneq S'$).
        By Definition~\ref{def:colourNeighbourhooh} we have $\gamma(u) = \gamma(w), \forall w \in S$ and there exists $v \in S$ such that $u \sim v$.
        Let us call $e = \{u, v\}$.
        $\forall w \in S$, there exists a path $\Gamma_{w, v}$ which is entirely contained in $S$ and which can be extended to $v$ by $e$, hence forming a path $\Gamma_{w, u}$ entirely contained in $S'$ .
        It follows that $S'$ is a local colour cluster such that $S \subsetneq S'$ \absurd.
    \end{description}
   
\end{proof}

The proof of Proposition~\ref{prop:colourComponent} highlights a constructive way to evaluate a colour component from one of its vertices $v \in V$ and directly yields a simple exploration procedure, as described in Algorithm~\ref{algo:colourComponent}.
Hence, the colour partition of a graph can be determined by means of a simple yet effective procedure, as described in Algorithm~\ref{algo:colourPartition}.

\begin{algorithm}[tp]
    \caption{\texttt{EvalColourComponent$(v, G)\algoout S$}}
    \label{algo:colourComponent}
    \KwIn{A vertex $v$ of the vertex-coloured graph $G=(V, E, \gamma)$}\\
    \KwOut{The colour component $S$ where $v$ belongs to}
    \begin{algorithmic}[1]
        \State $S \lassign \emptyset$;
        \State $N \lassign \{v\}$;
        \Repeat
            \State$S \lassign S \cup N$;
            \State$N \lassign N_\gamma(S)$;
        \Until{$N = \emptyset$;}
        \State\Return $S$;
    \end{algorithmic}
\end{algorithm}

\begin{algorithm}[tp]
    \caption{\texttt{EvalColourPartition$(G)\algoout \SSS^\star$}}
    \label{algo:colourPartition}
    \KwIn{A vertex-coloured graph $G=(V, E, \gamma)$}\\
    \KwOut{The colour partition $\SSS^\star = \CCC_\gamma(G)$}

    \begin{algorithmic}[1]
        \State $\SSS^\star \lassign \emptyset$;
        \State $K \lassign V$;
        
        \While{$K \ne \emptyset$}
            \State Let $v$ be a vertex from $K$;
            \State $S \lassign \texttt{EvalColourComponent}(v, G)$;
            \State $\SSS^\star \lassign \SSS^\star \cup \{S\}$;
            \State $K \lassign K\backslash S$;
        \EndWhile
        \State \Return $\SSS^\star$;
    \end{algorithmic}
\end{algorithm}

\subsection{The \texorpdfstring{$\gamma$}{gamma}-contraction operator}

With the concept of colour cluster, it is a natural consequence of Proposition~\ref{prop:commasscont} to claim the following.
\begin{proposition}[Colour contraction, variadic form]
    \label{prop:contractionVariadic}
    The colour-preserving contraction operator $\mmmm_G^\gamma$ admits a variadic form on colour clusters.
    In other words, given $U \subseteq V$ a colour cluster, then $G/_\gamma U$ is well defined and equips the resulting graph $G/U$ with the colouring function defined in \eqref{eg:colour-preserve}.
\end{proposition}
\begin{proof}
    The proposition directly follows from the fact that colour-preserving contraction is commutative and associative, analogously to regular graph contraction (cf.\ Section~\ref{ssec:graph-contraction}).
\end{proof}

Then, provided the definition of colour partition, under the variadic form of $\mmmm_G^\gamma$, we are now ready to refine Definition~\ref{def:colourContraction_v1} as follows.
\begin{definition}[$\gamma$-contraction (variadic form)]\label{def:colourContraction_v2}
    Let $G = (V, E, \gamma)$ be a vertex-coloured graph and let $\SSS^\star$ be its colour partition $\CCC_\gamma(G)$, made of $n'$ colour components $S_1, S_2, \dots, S_{n'}$.
    We define the \emph{$\gamma$-contraction} of $G$ -- and we denote it as $G/\gamma$ -- as the graph obtained by applying $G/_\gamma S$, $\forall S \in \SSS^\star$, \ie $G/_\gamma S_1/_\gamma\dots/_\gamma S_{n'}$.
\end{definition}

The latter definition further highlights that $\gamma$-contraction builds a novel graph $G' = (V', E', \gamma') = G/\gamma$ of order $n'$, where each node $v_i \in V'$ is the contraction of a colour component $S_i \in \CCC_\gamma(G)$ and it is equipped with the colour $\gamma(S_i)$.
It follows that $|G'| = n' = |\CCC_\gamma(G)| = |\CCC_{\gamma'}(G')|$, where the last equality follows from the maximality of $S_i$, \ie $\forall u', v' \in V'$ if $u' \sim v'$ then $\gamma'(u') \ne \gamma'(v')$.
In other words, $G'$ is a $c$-partite graph, where the partition is induced by $\gamma'$ (and $\gamma'$ is a proper colouring of $G'$).
Hence, we can use Algorithm~\ref{algo:colourPartition} to define a simple, intuitive way to perform $\gamma$-contraction, as resumed in Algorithm~\ref{algo:simpleColourContraction}.

\begin{algorithm}[tp]
    \caption{\texttt{simpleGammaContraction$(G)\algoout G'$}}
    \label{algo:simpleColourContraction}

    \KwIn{A vertex-coloured graph $G=(V, E, \gamma)$}\\
    \KwOut{The corresponding $\gamma$-contraction $G' = G/\gamma$}

    \begin{algorithmic}[1]
        \State $\SSS^\star \lassign \texttt{EvalColourPartition}(G)$;
        
        \ForAll{$S \in \SSS^\star$}
            \State $G \lassign G/_\gamma S$;
        \EndFor
        \State \Return $G$;
    \end{algorithmic}
\end{algorithm}

In particular, in the upcoming Section~\ref{sec:beta-contraction}, we explain how to leverage the variadic nature of $\mmmm_G^\gamma$ to build a novel local operator under the name of $\beta$-contraction, yielding an iterative form of $G/\gamma$ that we later use in Section~\ref{sec:algorithmic-framework} to build a fast local contraction algorithm replacing Algorithm~\ref{algo:simpleColourContraction}.


\section{The \texorpdfstring{$\beta$}{beta}-contraction operator}\label{sec:beta-contraction}

In this section, we introduce a novel contraction operator, called \emph{$\beta$-contraction}, which provides a canonical, locally defined refinement of $\gamma$-contraction for vertex-coloured graphs.
Rather than identifying all colour components at once, $\beta$-contraction collapses monochromatic regions according to a deterministic local rule, yielding a colour-preserving contraction that can be iterated until the full $\gamma$-contraction is recovered.

Repeated application of this operator induces a strictly decreasing sequence of graphs that converges to $G/\gamma$.
The constructive nature of $\beta$-contraction also yields an efficient evaluation strategy, which we describe alongside the formal development.

\subsection{Structural overview of the \texorpdfstring{$\beta$}{beta}-contraction operator}\label{ssec:beta-overview}

The definition of $\beta$-contraction exploits the variadic nature of colour-preserving contraction introduced in Section~\ref{sec:colour-contraction}.
Instead of contracting entire colour components, $\beta$-contraction identifies a colour sub-partition $\SSS_\beta$ whose elements are monochromatic connected subgraphs, not necessarily maximal.
Each set $S \in \SSS_\beta$ is then contracted into a single vertex, yielding a reduced graph $G/\beta$.

Formally, $\beta$-contraction is defined by a contraction mapping
\[
    \beta : V \to V' \cong \SSS_\beta,
\]
where the fibres of $\beta$ coincide with the elements of $\SSS_\beta$.
The essential requirement is that $\SSS_\beta$ refines the colour partition of $G$ and is itself induced by local adjacency relations between vertices of the same colour.

Clearly, the operator $\beta$ is not intended to coincide with $\gamma$-contraction in a single step.
Rather, it provides a canonical intermediate contraction whose repeated application yields $G/\gamma$.
In particular, only a few iterations are typically required to build $G/\gamma$ (it was shown in \cite{Lombardi_Onofri_22A} that $\sim 4$ iterations are required for $n\sim5\pow{}{4}$ and $m>n\log n$ random graphs), actually upper bounded by $\log_\varphi(n)$ (where $\varphi$ is the golden ratio, see later Theorem~\ref{theo:beta-convergence}), even if the worst case logarithmic number of iterations is achievable only on ad-hoc pathological input graphs.
The remainder of this section formalises the construction of $\beta$, proves its structural properties, and shows how it induces a well-defined contraction operator on coloured graphs.

We characterise the $\beta$-contraction operator by separating its definition into two logically distinct components, which are applied iteratively until $\gamma$-contraction is reached:
(i) the identification of a colour sub-partition $\SSS_\beta$, and
(ii) the contraction induced by that sub-partition.
Indeed, $\beta$-contraction itself is defined as a single contraction mapping; iteration only arises when $\beta$ is repeatedly evaluated and applied in order to recover $G/\gamma$.

More formally, let $G = (V, E, \gamma)$ be a vertex-coloured graph, let $\SSS_\beta$ be the colour sub-partition associated with $\beta$, and let $G' = (V', E', \gamma')$ be the graph resulting from the contraction procedure over $\beta$ (\ie obtained by contracting each $S \in \SSS_\beta$ into a single vertex).
The two components previously introduced can be described as follows.
\begin{description}
    \item[Identification of the contraction mapping]
    We define a contraction mapping $\beta : V \to V' \cong \SSS_\beta$ whose fibres coincide with the elements of the colour sub-partition.
    Each vertex $v \in V$ is mapped to the unique element of $\SSS_\beta$ containing it, and hence to the corresponding vertex of $V'$.
    The colour sub-partition $\SSS_\beta$ is not assumed \emph{a priori}, but is canonically induced by local adjacency relations between vertices sharing the same colour.
    Its construction relies exclusively on colour neighbourhoods and does not depend on global properties of the graph.

    \item[Induced contraction]
    Given the mapping $\beta$, the contracted graph $G' = G/\beta$ is uniquely determined.
    Since $\beta$ is colour-preserving, each vertex of $G'$ inherits the colour of the corresponding cluster.
    Contraction may be interpreted equivalently as the simultaneous application of colour-preserving contraction to all elements of $\SSS_\beta$, a viewpoint justified by the associativity of the contraction operator.
\end{description}
While this simultaneous contraction can be interpreted operationally as a ``multi-contraction'', no new graph-theoretic operation is introduced.
The resulting graph depends solely on the fibres of $\beta$ and coincides with the graph obtained by contracting each $S \in \SSS_\beta$ in any order, \ie considering $G/\beta$, where $\beta$ is interpreted as a colour function.

\paragraph*{Induced contraction}

The contraction induced by $\beta$ admits several equivalent characterisations.
Since the colour-preserving contraction is associative, the graph $G' = G/\beta$ can be described directly in terms of the induced colour sub-partition $\SSS_\beta$ as
\begin{equation}
    \begin{split}
        V' =&\ \{w_i \mid S_i \in \SSS_\beta\} \\
        E' =&\ \big\{\{w_i, w_j\} \mid w_i, w_j \in V' \;,\; N(S_i) \cap S_j \neq \emptyset\big\} \\
        \gamma'(w_i) =&\ \gamma(S_i), \quad \forall S_i \in \SSS_\beta .
    \end{split}
\end{equation}
Equivalently, once the contraction mapping $\beta : V \to V'$ is fixed, $G'$ can be defined in terms of the fibres of $\beta$ as
\begin{equation}
    \begin{split}
        V' =&\ \{\beta(v) \mid v \in V\} \\
        E' =&\ \big\{\{\beta(u), \beta(v)\} \mid \{u, v\}\in E\big\} \\
        \gamma'(w') =&\ \gamma(\beta^{-1}(w')), \quad \forall w' \in V' .
    \end{split}
\end{equation}
where in the last equation it is important to notice that $\beta^{-1}(w')$ is a monochromatic connected set $S$ (\ie colour component), hence compatible with the notation $\gamma(S)$ introduced in Definition~\ref{def:colourClusterComponent}.

These formulations are equivalent and uniquely determine the contracted graph.

\paragraph*{Identification of the contraction mapping}

For what concerns the first component, the underlying idea is to identify monochromatic components by means of a purely local rule.
To this end, each vertex $v \in V$ selects a representative vertex $\hat v$ within its colour neighbourhood, and this local choice is used to induce a global structure.
More precisely, the identification of the colour sub-partition $\SSS_\beta$ is based on a functional digraph $D$ naturally induced by a canonical choice rule (minimum for some ordering) on colour neighbourhoods.
Specifically, we associate with each vertex $v \in V$ a representative vertex
\[
\hat v \in N_\gamma(v) \cup \{v\},
\]
and encode this association as a directed edge $(v,\hat v)$.
The resulting digraph captures the aggregation of vertices into colour clusters.

The construction of $\beta$ relies on a deterministic and locally computable choice function.
To make this choice canonical, we fix a total order on the vertex set $V$.
This is naturally done by considering the bijection $V = \{v_0, \dots, v_{n-1}\} \leftrightarrow \{0, \dots, n-1\}$; accordingly, we write $v_i \leq v_j$ whenever $i \leq j$.
For each vertex $v$, we then define its representative as the minimum element of its colour neighbourhood, choosing $v$ itself whenever no other vertex of the same colour is adjacent.

This construction induces a functional digraph whose structure encodes the desired colour sub-partition.
The following theorem formalises this intuition and characterises the resulting digraph.

\begin{theorem}[Creation and characterisation of $D$]\label{theo:D-forest}
    Let $G = (V, E, \gamma)$ be a vertex-coloured graph where vertices are enumerated in $\{0, \dots, n-1\}$.
    Define the digraph with self-loops $D = (V, B)$, where
    \begin{equation}\label{eq:B}
        B = \{(v, \hat v) \mid v \in V\},
    \end{equation}
    and
    \begin{equation}\label{eq:hatv}
        \hat v = \min\big(N_\gamma(v) \cup \{v\}\big), \qquad \forall v \in V.
    \end{equation}
    Then $D$ is a forest.
    
    In particular, each connected component $T \in \CCC(D)$ is a tree whose vertex set $V_T$ is a monochromatic connected subgraph of $G$.
    Consequently, the family $\CCC(D)$ induces a colour sub-partition $\SSS_\beta$ of $G$.

    Moreover, for each tree $T \in \CCC(D)$, the vertex
    \[
        r_T = \min\{v \in V_T\}
    \]
    satisfies the property that, for every $v \in V_T$, the unique path $\Gamma_{v,r_T}$ is directed and maximal with respect to $v$ (\ie, the longest directed path with source in $v$).
    Hence $r_T$ provides a natural root for $T$ and a canonical representative for the corresponding colour cluster $S = V_T \in \SSS_\beta$.
    In particular, the set of roots $\{r_T\}_{T \in \CCC(D)}$ induces a natural enumeration of the clusters in $\SSS_\beta$.
\end{theorem}
\begin{proof}
    We prove the statement by addressing each claim individually
    \begin{enumerate}
        \item
        Let us prove that $D$ is a forest.
        We recall that, in order for $D$ to be a forest, then given two distinct nodes $u, v \in V$ there exists at most a single path $\Gamma_{u, v}$ joining them.
    
        Without loss of generality, let $u > v$ and assume $u, v$ are on the same connected component, hence $\exists \Gamma_{u, v} = (w_0, \dots, w_\ell)$ a path connecting them.
        We now prove that $\Gamma_{u, v}$ is unique.
        
        In particular, from \eqref{eq:B}, each node $w \in V$ has $\degree_D^\scout(w) = 1$, and we have $v < u, \forall u \in N^\scin(v)$ and $u \geq v$ for the single node $u \in N^\scout(v)$ (where equality holds only if $u=v$, \ie v has a self loop).
        It follows that any undirected path in $D$ can be decomposed into at most two monotone directed segments, \ie (since $u > v$) $\exists a \in \{1, \dots, \ell\}$ such that $w_0 > w_1 > \dots w_a$ and (unless $a = \ell$) $w_a < w_{a+1} < \dots w_\ell$.
        In fact, if more than a change in monotony exists, then there exists a node $w_{\hat {i}}$ such that $w_{\hat{i}-1}<w_{\hat {i}}$ and $w_{\hat {i}}> w_{\hat{i}+1}$, which is impossible since it would imply that $\degree^\scout(v) = 2$.
    
        In particular, the two monotone paths $\Gamma_{u, w_a}$ and $\Gamma_{v, w_a}$ are unique according to the same argument, hence $\Gamma_{u, v}$ is unique as well.

        \item
        Let us prove that a connected component $T \in \CCC(D)$ spans a colour cluster.
        Being $D$ a forest, it is straightforward that $T$ is a maximal tree in $D$.
        In particular, it follows from \eqref{eq:hatv} that $\forall u, v \in V_T$, then $\gamma(u) = \gamma(v)$, hence $\langle T \rangle_G$ is a (connected, colour preserving) sub-graph of $G$ whose vertices (\ie $V_T$) form a colour cluster.
        
        Clearly, being $\CCC(D)$ a partition, we have that $T_i$ and $T_j$ are disjoint $\forall T_i, T_j \in \CCC(D)$, hence
		\begin{equation}\label{eq:Sbeta}
        	\SSS_\beta = \SSS = \{V_T \mid T \in \CCC(D)\}
		\end{equation}
		is a colour subpartition.

        \item
        Let us prove that for any $T \in \CCC(D)$, given $r_T = \min\{v \in V_T\}$, the unique path $\Gamma_{v, r_T}$ is directed $\forall v \in V_T$ and it is maximal \wrt $v$.
        Since $\degree_D^\scout(u) = 1, \forall u \in V_T$, it is easy to see that, $\forall v \in V_T$, the longest directed path $\Gamma_{v, *}$ starting from $v$ is uniquely determined by \eqref{eq:hatv} and, in particular, it is monotone.
        Therefore, $\Gamma_{v, *}$ ends in a node $r_v = \min\{z \in \Gamma_{v, *}\}$.
        It follows from (1.) that $\forall u, v \in V_T, \exists w \in \Gamma_{u, v}$ such that the path can be split into two monotone directed paths $\Gamma_{u, w}$ and $\Gamma_{v, w}$, hence $w = \min\{z \in \Gamma_{u, v}\}$ and $w \in \Gamma_{u,*} \cap \Gamma_{v, *}$.
        In particular, $r_v = r_w$ and $r_u = r_w$, hence $r_u = r_v, \forall u, v \in V_T$, so $r_v = r_T$ holds for all $v \in V_T$.
                
        The vertex $r_T$ is then a natural choice for being the root of the tree $T$.
        Furthermore, if we let $R = \{r_T, \forall T \in \CCC(D)\}$ be the set of the roots within the forest, then it is natural to consider the map $\tilde\pi$:
        \begin{equation}\label{eq:pitilde}
        	\begin{matrix}
				\tilde\pi & : & \SSS & \to & R &\subseteq&V\\
				&&V_T & \mapsto & r_T&&
			\end{matrix}
		\end{equation}
        or, analogously, the map $\hat\pi : V \to R\subseteq V$ defined as
        \begin{equation}\label{eq:pihat}
        	\hat\pi(v) = \tilde\pi(S), \quad \forall v \in S, \quad \forall S \in \SSS\ .
        \end{equation}
        
        \item
        We finally show how the roots of the forest induce a natural enumeration of the colour clusters $S \subseteq \SSS$.
        Let $n' = |\SSS|$, then we can build the natural bijection $\hat\alpha : R \to \{0, \dots, n'-1\}$ that preserves the ordering on $R$, \ie
        \begin{equation}\label{eq:alphahat}
        	\hat\alpha(r_T) \leq \alpha(r_{T'}) \quad \iff \quad r_T \leq r_{T'}, \qquad \forall r_T, r_{T'} \in R\ ,
        \end{equation}
        where equality holds if and only if $T=T'$.
        Hence, $\hat\alpha$ induces an enumeration $\tilde\alpha$ on $\SSS$ via $\tilde\pi$, \ie
        \begin{equation}\label{eq:alphatilde}
        	\begin{matrix}
				\tilde\alpha & : & \SSS & \to & \{0, \dots, n'-1\}\\
				&& S & \mapsto & \hat\alpha(\tilde\pi(S))
			\end{matrix}\ .
        \end{equation}
    \end{enumerate}
    The points (1.)--(4.) conclude the proof.
\end{proof}
\begin{remark}[On the constructiveness of the proof of Theorem~\ref{theo:D-forest}]\label{rm:theo-D-forest}
	In the proof of Theorem~\ref{theo:D-forest}  we have constructively defined the colour sub-partition $\SSS_\beta$ in \eqref{eq:Sbeta} and we have introduced two couples of analogous maps, whose form we recall in the following for ease of notation:
	\[
		\begin{matrix}
			\hat\pi:V\to R, \text{ in \eqref{eq:pihat}}
			& \qquad &
			\hat\alpha : R \to \{0, \dots, n'-1\}, \text{ in \eqref{eq:alphahat}}
			\\
			\tilde\pi:\SSS \to R, \text{ in \eqref{eq:pitilde}}
			& &
			\tilde\alpha : \SSS \to \{0, \dots, n'-1\}, \text{ in \eqref{eq:alphatilde}}
		\end{matrix}\ .
	\]
	In particular, the formulation $\tilde \alpha$ and $\tilde\pi$ suffice to establish Theorem~\ref{theo:D-forest} , while $\hat\alpha$ and $\hat\pi$ are used in the following to build $\beta$.
\end{remark}

The identification of the contraction mapping $\beta$ proceeds in four steps, following naturally from \remarkname~\ref{rm:theo-D-forest}:
\begin{description}
    \item[Creation of $D$]
    We create a digraph with self loops $D = (V, B)$ with $B$ defined as in \eqref{eq:B}.
    In particular, $D$ is a rooted forest (with root set $R$) whose components in $\CCC(D)$ define the colour sub-partition $\SSS_\beta$.
    
    \item[Creation of $\pi : V \to R$]
	The map $\pi$ is already defined as $\hat\pi$ \eqref{eq:pihat}.
    
    \item[Creation of $\alpha : R \to V'$]
    Let us consider a set of novel vertices $V' = \{w_i, \forall i \in \{0, \dots, n'-1\}\}$ enumerated in $\{0, \dots, n'-1\}$, where $n' = |\SSS_\beta|$.
    Since clusters $S \in \SSS_\beta$ are enumerated in the same interval according $\hat\alpha$, as defined in \eqref{eq:alphahat}, hence it is natural to build the bijection $\alpha$ between roots and novel vertices as follows:
    \begin{equation}
    	\begin{matrix}
    		\alpha & : & R & \to & V'\\
			&& r & \mapsto & w_{\tilde\alpha(v)}
    	\end{matrix}
    \end{equation}

    \item[Creation of $\beta$]
    The map $\beta : V \to V'$ directly follows as the composition $\alpha\circ\pi$, \ie
    \begin{equation}\label{eq:betadef}
        \begin{matrix}
    		\beta & : & V & \to & V'\\
			&& v & \mapsto & \alpha(\pi(v))
    	\end{matrix}\ .
    \end{equation}
\end{description}

An example of the described procedure can be found in Figure~\ref{fig:contraction-example}.

\begin{remark}[$D$ identifies colour clusters that might be not maximal]\label{rm:caso-patologico}
    It is important to notice that $S \in \SSS_\beta$, despite being a colour cluster, might or might be not a colour component depending on the chosen enumeration of vertices, \ie.
    A simple counter-example is given by the path graph $G = P^4 = (\{v_0, \dots v_3\}, E, \gamma)$ where $E = \{\{v_0,v_2\},\{v_1,v_3\},\{v_2,v_3\}\}$ and $\gamma(v) : V \to \{c\}$ which is represented in Figure~\ref{fig:caso-patologico}.
    Here, we have $\hat v_0 = v_0$, $\hat v_1 = v_1$, $\hat v_2 = v_0$, and $\hat v_3 = v_1$, hence resulting in $D = (V, \{(v_2, v_0), (v_3, v_1)\})$ which identifies the two trees with vertex sets $S_0 = \{v_0, v_2\}$ and $S_1 = \{v_1, v_3\}$.
\end{remark}

The behaviour introduced in \remarkname~\ref{rm:caso-patologico} motivates the iterative application of $\beta$-contraction, which progressively merges such sub-clusters until full $\gamma$-contraction is achieved.

\begin{figure}[tp]
    \centering
    \centering
    \begin{subfigure}{.6\linewidth}
        \centering
        \scalebox{.75}{
%
%
\begin{tikzpicture}[
	scale=.75,
	v/.style={draw,circle,minimum size=.9cm},
	Econn/.style={-},
	Bconn/.style={->, ultra thick},
	pimap/.style={->, dashed},
	betam/.style={->, thick, dashed},
]

    \def\dx{2.5}
    \def\kk{0.866025}
    
	\node[v, fill=C3!80] (20) at (-\dx*\kk*3.0, +\dx*0.5) {$20$};
	\node[v, fill=C3!80] (19) at (-\dx*\kk*3.0, -\dx*0.5) {$19$};

    \node[v, fill=C1!80] (22) at (-\dx*\kk*2.0, +\dx*2.0) {$22$};
	\node[v, fill=C3!80] (21) at (-\dx*\kk*2.0, +\dx*1.0) {$21$};
	\node[v, fill=C2!80] (18) at (-\dx*\kk*2.0, -\dx*1.0) {$18$};
	\node[v, fill=C2!80] (17) at (-\dx*\kk*2.0, -\dx*2.0) {$17$};
    
	\node[v, fill=C2!80] (23) at (-\dx*\kk*1.0, +\dx*2.5) {$23$};
	\node[v, fill=C1!80] (05) at (-\dx*\kk*1.0, +\dx*0.5)  {$5$};
	\node[v, fill=C2!80] (04) at (-\dx*\kk*1.0, -\dx*0.5)  {$4$};
	\node[v, fill=C2!80] (16) at (-\dx*\kk*1.0, -\dx*2.5) {$16$};
    
	\node[v, fill=C2!80] (06) at (+\dx*\kk*0.0, +\dx*2.0)  {$6$};
	\node[v, fill=C2!80] (00) at (+\dx*\kk*0.0, +\dx*1.0)  {$0$};
	\node[v, fill=C1!80] (03) at (+\dx*\kk*0.0, -\dx*1.0)  {$3$};
	\node[v, fill=C2!80] (15) at (+\dx*\kk*0.0, -\dx*2.0) {$15$};
	
	\node[v, fill=C3!80] (07) at (+\dx*\kk*1.0, +\dx*2.5)  {$7$};
    \node[v, fill=C2!80] (01) at (+\dx*\kk*1.0, +\dx*0.5)  {$1$};
	\node[v, fill=C2!80] (02) at (+\dx*\kk*1.0, -\dx*0.5)  {$2$};
	\node[v, fill=C2!80] (14) at (+\dx*\kk*1.0, -\dx*2.5) {$14$};

    \node[v, fill=C3!80] (08) at (+\dx*\kk*2.0, +\dx*2.0)  {$8$};
	\node[v, fill=C3!80] (09) at (+\dx*\kk*2.0, +\dx*1.0)  {$9$};
	\node[v, fill=C1!80] (12) at (+\dx*\kk*2.0, -\dx*1.0) {$12$};
	\node[v, fill=C2!80] (13) at (+\dx*\kk*2.0, -\dx*2.0) {$13$};
    
	\node[v, fill=C1!80] (10) at (+\dx*\kk*3.0, +\dx*0.5) {$10$};
	\node[v, fill=C1!80] (11) at (+\dx*\kk*3.0, -\dx*0.5) {$11$};
	
	\draw
        (00) edge[Econn] (01)
        (00) edge[Econn] (06)
        (01) edge[Econn] (02)
        (02) edge[Econn] (03)
        (02) edge[Econn] (12)
        (03) edge[Econn] (04)
        (03) edge[Econn] (15)
        (04) edge[Econn] (16)
        (04) edge[Econn] (18)
        (05) edge[Econn] (21)
        (05) edge[Econn] (22)
        (06) edge[Econn] (07)
        (06) edge[Econn] (23)
        (07) edge[Econn] (08)
        (08) edge[Econn] (09)
        (09) edge[Econn] (10)
        (10) edge[Econn] (11)
        (11) edge[Econn] (12)
        (13) edge[Econn] (14)
        (14) edge[Econn] (15)
        (16) edge[Econn] (17)
        (17) edge[Econn] (18)
        (18) edge[Econn] (19)
        (19) edge[Econn] (20)
        (20) edge[Econn] (21)
        (21) edge[Econn] (22)
        (22) edge[Econn] (23)
	;
	
	\draw
        (00) edge[pimap, out=+30, in=+60, looseness=8] node[above] {0} (00)
        (01) edge[pimap, bend left ]          (00)
        (02) edge[pimap, bend left ]          (00)
        (03) edge[pimap, loop above] node {1} (03)
        (04) edge[pimap, out=+150, in=+120, looseness=8] node[above] {2} (04)
        (05) edge[pimap, out=-60, in=-30, looseness=8] node[below] {3} (05)
        (06) edge[pimap, bend right]          (00)
        (07) edge[pimap, loop left ] node {4} (07)
        (08) edge[pimap, bend left ]          (07)
        (09) edge[pimap, bend left ]          (07)
        (10) edge[pimap, loop left] node {5} (10)
        (11) edge[pimap, bend left ]          (10)
        (12) edge[pimap, bend left ]          (10)
        (13) edge[pimap, loop right] node {6} (13)
        (14) edge[pimap, bend left ]          (13)
        (15) edge[pimap, bend left ]          (13)
        (16) edge[pimap, bend right]          (04)
        (17) edge[pimap, bend right]          (04)
        (18) edge[pimap, bend right]          (04)
        (19) edge[pimap, loop below] node {7} (19)
        (20) edge[pimap, bend left ]          (19)
        (21) edge[pimap, bend left ]          (19)
        (22) edge[pimap, bend left ]          (05)
        (23) edge[pimap, bend right]          (00)
    ;
    
	\draw
		(01) edge[Bconn] (00)
		(02) edge[Bconn] (01)
		(06) edge[Bconn] (00)
		(08) edge[Bconn] (07)
		(09) edge[Bconn] (08)
		(11) edge[Bconn] (10)
		(12) edge[Bconn] (11)
		(14) edge[Bconn] (13)
		(15) edge[Bconn] (14)
		(16) edge[Bconn] (04)
		(17) edge[Bconn] (16)
		(18) edge[Bconn] (04)
		(20) edge[Bconn] (19)
		(21) edge[Bconn] (20)
		(22) edge[Bconn] (05)
		(23) edge[Bconn] (06)
	;
	
\end{tikzpicture}}
        \caption{}
        \label{fig:contraction-example-a}
    \end{subfigure}
    \hfill
    \begin{subfigure}{.3\linewidth}
        \centering
        \scalebox{.8}{
%
%
\begin{tikzpicture}[
	scale=.75,
	v/.style={draw,circle,minimum size=.9cm},
	Econn/.style={-},
	Bconn/.style={->, ultra thick},
	pimap/.style={->, dashed},
	betam/.style={->, thick, dashed},
]

    \def\dx{2.5}
    
	\node[v, fill=C2!80] (0) at (1*\dx, 2*\dx) {$0$};
	\node[v, fill=C1!80] (1) at (1*\dx, 0*\dx) {$1$};
	\node[v, fill=C2!80] (2) at (0*\dx, 0*\dx) {$2$};
	\node[v, fill=C1!80] (3) at (0*\dx, 2*\dx) {$3$};
	\node[v, fill=C3!80] (4) at (2*\dx, 2*\dx) {$4$};
	\node[v, fill=C1!80] (5) at (2*\dx, 1*\dx) {$5$};
	\node[v, fill=C2!80] (6) at (2*\dx, 0*\dx) {$6$};
	\node[v, fill=C3!80] (7) at (0*\dx, 1*\dx) {$7$};
    \node at (0, -3.8) {\phantom{x}};
	
	\draw
        (0) edge[Econn] (1)
        (0) edge[Econn] (3)
        (0) edge[Econn] (4)
        (0) edge[Econn] (5)
        (1) edge[Econn] (2)
        (1) edge[Econn] (6)
        (2) edge[Econn] (7)
        (3) edge[Econn] (7)
        (4) edge[Econn] (5)
	;
	
\end{tikzpicture}}
        \caption{}
        \label{fig:contraction-example-b}
    \end{subfigure}
    \caption[Example of 1-iteration $\beta$-contraction]{
        Example of a 1-iteration $\beta$-contraction of the graph $G$ to the graph $G' = G/\gamma$.
        \subref{fig:contraction-example-a}
        The graph $G = (V, E)$.
        The directed edges $B$ of the digraph $D = (V, B)$ are highlighted as bold arrows over $G$ (self-loops are omitted).
        The map $\pi$ is represented with dashed bend arrows which link each node within a colour cluster to the corresponding tree root \ie nodes where $\pi$ form a self-loop.
        Numeric values of the map $\hat\alpha$ are reported on the $\pi$ self-loops.
        The map $\beta$ is not reported for the sake of readability (see Figure~\ref{fig:caso-patologico}).
        \subref{fig:contraction-example-b}
        The contracted graph $G' = (V', E') = G/\beta = G/\gamma$.
        Vertices in $V'$ (\ie the codomain of $\beta$) are enumerated according to values on self-loops from $\pi$.
    }
    \label{fig:contraction-example}
\end{figure}

\begin{figure}[tp]
    \centering
    \def\scalefactor{.9}
    \scalebox{\scalefactor}{    \begin{tikzpicture}[
        tn/.style={draw,circle,fill=C1!80},
        Econn/.style={-},
        Bconn/.style={->, ultra thick},
        pimap/.style={->, dashed},
        betam/.style={->, thick, dotted},
    ]
        \def\dx{2}
    
        \node[tn] (0) at (0*\dx, 0) {0};
        \node[tn] (2) at (1*\dx, 0) {2};
        \node[tn] (3) at (2*\dx, 0) {3};
        \node[tn] (1) at (3*\dx, 0) {1};
        \node[tn] (a) at (1*\dx,-2) {0};
        \node[tn] (b) at (2*\dx,-2) {1};
        \draw[Econn] (0) -- (2) -- (3) -- (1);
        \draw[Bconn] (2) -- (0);
        \draw[Bconn] (3) -- (1);
        \path (0) edge [pimap, loop left] node{0} (0);
        \path (2) edge [pimap, bend right] (0);
        \path (3) edge [pimap, bend left ] (1);
        \path (1) edge [pimap, loop right] node{1} (1);
        \draw[betam] (0) -- (a);
        \draw[betam] (2) -- (a);
        \draw[betam] (3) -- (b);
        \draw[betam] (1) -- (b);
        \draw[Econn] (a) -- (b);

        \node[] at (-\dx, 0) {$G$};
        \node[] at (-\dx,-2) {$G'$};
        \node[] at (4*\dx, 0) {\phantom{$G'$}};
    \end{tikzpicture}}
    \caption[A graph sample where clusters are not component]{
        A small example of a graph $G = (V, E)$ for which $\beta$ determines two distinct colour clusters instead of a single colour component.
        The corresponding contracted graph $G' = G/\beta = (V', E')$ is also reported.
        The directed edges $B$ of the digraph $D = (V, B)$ are highlighted as bold arrows over $G$ (self-loops are omitted).
        The map $\pi$ is represented with dashed bend arrows and the numeric values of the map $\hat\alpha$ are reported on the $\pi$ self-loops.
        The map $\beta$ is represented as dotted arrows connecting $G$ with $G'$.
    }
    \label{fig:caso-patologico}
\end{figure}

\subsection{Convergence of the \texorpdfstring{$\beta$}{beta}-contraction operator}\label{subsec:beta-convergence}

The contraction mapping $\beta$ introduced in Section~\ref{ssec:beta-overview} is defined by a purely local choice rule on colour neighbourhoods.
As discussed in Remark~\ref{rm:caso-patologico}, a single application of $\beta$ does not, in general, identify maximal colour components, but rather produces a refinement of the colour partition which depends on the chosen enumeration of the vertices.
It is therefore natural to investigate the behaviour of $\beta$ under progressive iteration.

Given a coloured graph $G = (V, E, \gamma)$, we consider the sequence of graphs
\[
    G_0 = G, 
    \qquad 
    G_{k+1} = G_k / \beta_k ,
\]
where $\beta_k$ denotes the contraction mapping constructed on $G_k$ according to \eqref{eq:betadef}.
The fundamental questions are whether this process always terminates, whether it converges to the canonical colour contraction $G/\gamma$, and how many iterations are required in the worst case.

Since $\beta$ is colour preserving by construction, vertices of distinct colours are never identified.
As a consequence, distinct colour components evolve independently under iteration.
More precisely, if $\{, \dots, C_m\}$ denotes the set of colour components of $G$, then the restriction of $\beta$ to each induced subgraph $\langle C_i \rangle_G$ defines an independent contraction process, and the colour components of $G_k$ are exactly given by the $k$-th iteration of $\beta$-contraction to $\langle C_i \rangle_G$, for every iteration $k$.

Therefore, all global convergence properties of the $\beta$-contraction reduce to the analysis of a single colour component.
In the remainder of this section, we shall thus assume without loss of generality that $G$ is connected and all vertices share the same colour, \ie we consider the constant colouring function $\gamma_\textit{id}$ defined as follows:
\begin{equation}
    \begin{matrix}
        \gamma_\textit{id} : & V & \to & C \\
            & v & \mapsto & 0
    \end{matrix}\ .
\end{equation}

Under these assumptions, the canonical colour contraction $G/{\gamma_\textit{id}}$ consists of a single vertex.
Proving convergence of the iterative $\beta$-contraction is therefore equivalent to proving that repeated application of $\beta$ reduces $G$ to a singleton graph in a finite number of steps.

We are now ready to state the main convergence result for the $\beta$-contraction as follows.
\begin{theorem}[Convergence of the $\beta$-contraction of a colour component]\label{theo:beta-convergence}
	Let $G = (V, E, \gamma_\textit{id})$ be a connected coloured-vertex graph where vertices are enumerated in $\{0, \dots, n-1\}$.
    Iterated applications of $\beta$-contraction, defined according to \eqref{eq:betadef}, converge to a single vertex.
    Moreover, the number of iterations is bounded by
    \[
        \lfloor \log_\varphi(n) \rfloor ,
    \]
    where $\varphi = \frac{1+\sqrt5}{2}$ denotes the golden ratio.
\end{theorem}

In order to simplify the proof of Theorem~\ref{theo:beta-convergence}, we first prove that, (i) at any step, the roots of $D$ are pairwise non-adjacent and (ii) there can not be any vertex that forms an isolated component in $D$ for any two subsequent iterations. More formally, we have the following two lemmas:

\begin{lemma}[Characterisation of the roots of $D$]\label{lemma:charR}
    Let $G = (V, E, \gamma_\textit{id})$ be defined as in Theorem~\ref{theo:beta-convergence} and let $D = (V,B)$ be the oriented forest built as in Theorem~\ref{theo:D-forest} with roots set $R$.
    We have that $r_i \not\sim r_j, \forall r_i, r_j \in R$
\end{lemma}
\begin{proof}
    The result is an immediate consequence of the construction of $D$.
    Let us consider $u, v \in V$ such that $u \sim v$, and let $u < v$ without loss of generality.
    Then, according \eqref{eq:hatv}, either $(v, u) \in B$ or $(v, w) \in B$ for some $w \in N(v)$ such that $w < v$.
    In both cases, $v \not \in R$, concluding the proof.
\end{proof}

\begin{lemma}[Characterisation of the isolated vertices of $D$]\label{lemma:charSingleTrees}
    Let $G = (V, E, \gamma_\textit{id})$ be a non-trivial graph ($|V| > 1$) defined as in Theorem~\ref{theo:beta-convergence} and let $D = (V,B)$ be the oriented forest built as in Theorem~\ref{theo:D-forest} with roots set $R$ and contraction mapping $\beta$.
    Let also $G' = (V', E', \gamma'_\textit{id})$ be the contracted graph $G/\beta$ and let $D' = (V', B')$ be its oriented forest with root set $R'$.
    We have that any isolated vertex in $D$ cannot be isolated in $D'$, \ie $\forall u \in V$ such that $\{u\} \in \CCC(D)$, if $u' = \beta(u)$, then $\{u'\} \not \in \CCC(D')$.
\end{lemma}
\begin{proof}
    The claim follows directly from the definition of $D$.
    Let $u \in V$ be an isolated vertex in $D$ with $|V|>1$.
    Do note that the requirements of $G$ connected and $|V|>1$ ensure that $N(u) \ne \emptyset$.
    It follows that, $\forall v \in N(u), \exists w \in N(v)$ such that $w < u$.
    If $r$ is the root associated with $w$ (and $v$), we have that $u > r$ and, since $u \in R$, it follows from \eqref{eq:alphahat} that hence $\beta(u) < \beta(r)$.
    Finally, let $u' = \beta(u)$ and $v' = \beta(v) = \beta(w) = \beta(r)$.
    We can conclude the proof by observing that, since $(u', v') \in E'$, we have $(u', v') \in B'$, \ie $u'$ is not isolated in $D'$.
\end{proof}

\begin{remark}[On the hypotheses of Lemma~\ref{lemma:charSingleTrees}]
    Do note that if $|V| = 1$ there is nothing to prove, since $D$ is already an isolated vertex and hence $G = G'$ and $D'$ is an isolated vertex as well.
    This falsifies the thesis of the lemma, yet it is out of the scope of its applicability, since when $|V|=1$ then the graph can not be contracted anymore.
\end{remark}

Also, before discussing the proof of Theorem~\ref{theo:beta-convergence}, we provide a weaker version of it which is easier to prove but does not provide a strict complexity bound:

\begin{theorem}[Convergence of the $\beta$-contraction of a colour component (weak form)]\label{theo:beta-convergence-weak}
    Let $G = (V, E, \gamma_\textit{id})$ be a connected coloured-vertex graph where vertices are enumerated in $\{0, \dots, n-1\}$.
    Iterated applications of $\beta$-contraction, defined, where $\beta$ is defined according \eqref{eq:betadef}, converge to a single node and the number of iterations is upper bounded by $\log_{\sfrac32}(n)$.
    In particular, the worst-case number of iterations is lower bounded by $\log_2(n)$.
\end{theorem}

\begin{proof}
    We divide the proof into three steps for the reader's convenience.

    \noindent\textbf{- $\beta$-contraction converges to a single node.}
    We recall that a single iteration corresponds to evaluate the forest $D = (V, B)$ which implicitly defines the colouring $\beta$, and that nodes of $G/\beta$ are in bijection with the set $R$ of the roots of $D$ defined as $r_T = \min\{v \in V_T\}$, for all the trees $T \in \CCC(D)$ (cf.\ Theorem~\ref{theo:D-forest} and its proof).
    It is also important to notice that $|R| \leq |V|$, where clearly $|R| = |V|$ if and only if $R = V$, hence the convergence condition can be rephrased as $R = V$.
    
    If we collect all the iteration altogether, the contraction produces a sequence of graphs $G = G_0, G_1, \dots, G_k = G/\gamma$, with $G_i = (V_i, E_i)$, and a sequence of forests $D_0, \dots, D_k$, with $D_i = (V_i, B_i)$ and set of roots $R_i$, where $R_i \equiv V_{i+1}, \forall i \in\{0, \dots, k-1\}$ and $R_k = V_k$ being the stopping condition.

    Since we have that $|V_i| \geq |R_i| > 0, \forall i \in \{0, \dots, k\}$ and equality holds if and only if $i = k$, then it is clear that $|V_i|$ is a monotone non-increasing sequence and that $\beta$-contraction reaches convergence.
    Furthermore, $|V_k| = 1$; in fact, assuming $G_k$ connected and letting $v^* = \min\{v \in V_k\}$, if \adabsurdum $|V_k| > 1$ then there exists at least one $v \in N_{G_k}(v^*)$ $G_k$, and hence $\pi(v) = v^*$ and $R_k \ne V_k$.\absurd
    
    It follows that $\beta$-contraction converges to a single node, concluding the first part of the proof.

    \smallskip

    \noindent\textbf{- the worst-case number of iterations is lower bounded by $\log_2(n)$.}
    We now present a simple yet instructive bound from below for the worst case number of iterations.

    Let $n_i = |V_i| = |R_{i-1}|$, for $i \in \{1, \dots, k\}$ and let $n_0 = |V_0| = |V| = n$.
    Then, if at a given iteration $i$ we have no isolated vertices in $D_i$, then $|T| > 1, \forall T \in \CCC(D_i)$ (\ie, all the clusters have size at least two) and we can conclude that $n_{i+i} = |R_i| \leq \sfrac12 \ n_i$.
    If this would hold for all the iterations, then the convergence time would clearly be $\log_2(n)$, which is a lower bound for the worst-case scenario.
    It follows that understanding how many isolated vertices are present in $D_i$ is crucial to correctly estimate the total number of iterations.
    
    \smallskip
    
    \noindent\textbf{- $\log_{\sfrac32}(n)$ is an upper bound for the convergence time.}
    We now prove that $\beta$-contraction converges at least as fast as $\log_{\sfrac32}(n)$ iterations by collecting Lemma~\ref{lemma:charR} and~\ref{lemma:charSingleTrees}.
    This paves the way to provide the proof of the exact number of iterations.
    
    In particular, we notice that, given an iteration $i$, there can not be any two adjacent isolated vertices of $D_i$ due to Lemma~\ref{lemma:charR}.
    Hence, any isolated vertex of $D_i$ gets contracted at step $i+1$ (cf.\ Lemma~\ref{lemma:charSingleTrees}) with a vertex obtained at step $i$ from a colour cluster of size greater than one, worst case being of size 2.
    It is rather clear that the worst case at step $i+1$ is achieved when all the isolated vertices at step $i$ gets contracted at step $i+1$ in a minimal-sized cluster, \ie of size 2.
    This upper bounds the total number of isolated vertices at iteration $i$ to (at most) half of the total number of the colour clusters, with the other half made of clusters of size 2.
    Hence, it is easy to see that the case where $\sfrac23$ of the $n_i$ vertices form $\sfrac13\ n_i$ colour clusters of size 2 and the remaining $\sfrac13$ of the $n_i$ vertices are isolated (\ie forming $\sfrac13\ n_i$ colour clusters of size 1) provides an upper bound for the worst case at iteration $i$. This yields $n_{i+1} = \sfrac23\ n_i$ which, if assumed holding true for each iteration, provides an upper bound of the iteration required in the worst case of $\log_{\sfrac32}(n)$.
    However, as we see in the proof for Theorem~\ref{theo:beta-convergence} (see Appendix~\ref{app:fibonacci}), this bound is not strict, yet the methodology provides useful hints on how to obtain it
\end{proof}

\begin{figure}[t]
    \centering
    \includegraphics[width=.99\linewidth]{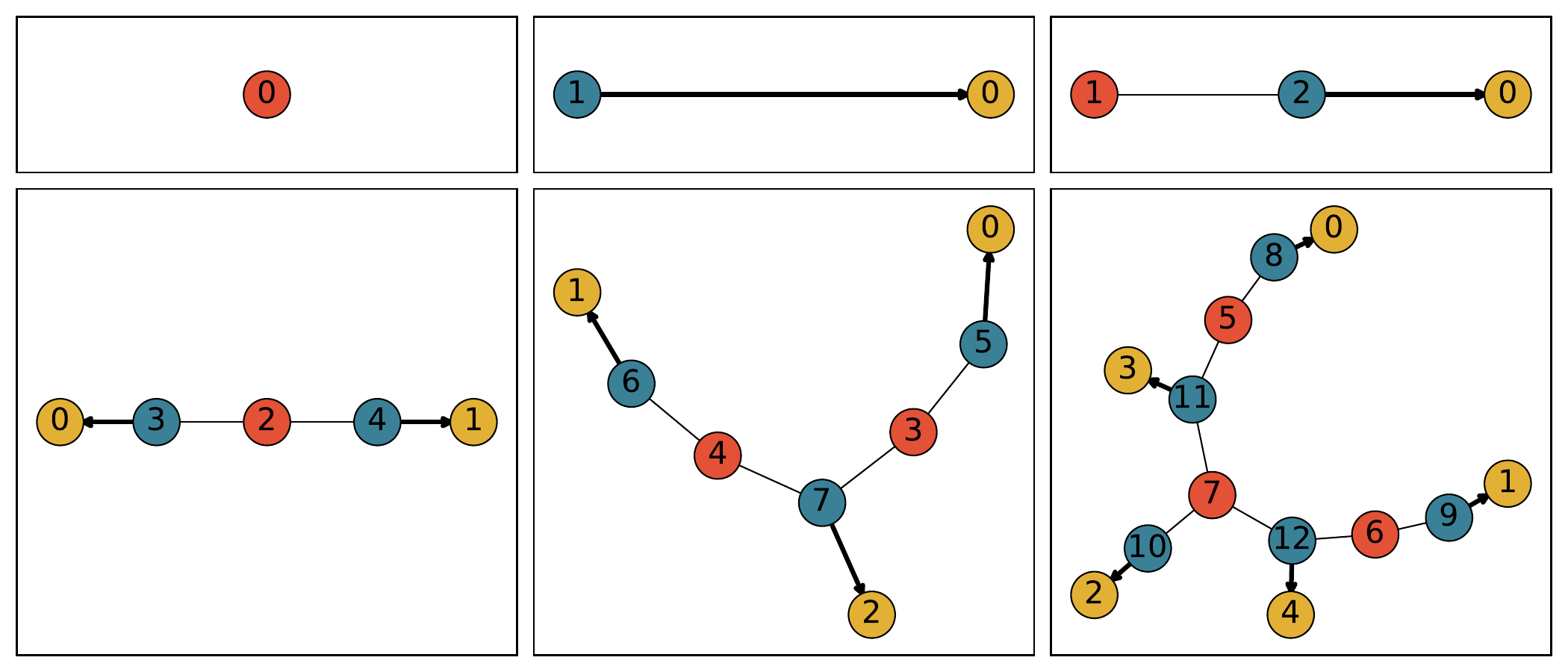}
    
    \begin{tikzpicture}[vertex/.style={draw, circle, black}]
        \node[vertex, fill=R_P_color, label={[black,align=left]right:$R\backslash P$}] () at (6, 0) {};
        \node[vertex, fill=Q_color, label={[black,align=left]right:$Q$}] () at (3, 0) {};
        \node[vertex, fill=P_color, label={[black,align=left]right:$P$}] () at (0, 0) {};
    \end{tikzpicture}
    \caption{
        First six iterations of the Fibonacci-type worst-case construction for $\beta$-contraction.
        Trees are either of order 1 or 2.
        Vertices are coloured according to their role in the functional digraph $D_i$, see also Appendix~\ref{app:fibonacci}:
        clusters of size 2 are divided into roots ($P$) and non-roots ($Q$); remaining nodes ($R\setminus P = V\setminus (Q \cup P)$) are roots of clusters of size 1.
        The construction enforces $G_{i+1}/\beta = G_i$ while minimising $|V(G_i)|$.
    }
    \label{fig:constructiveG-main}
\end{figure}

We are now ready to provide the exact convergence time in the worst-case scenario via the following:

\begin{proof}[Proof of Theorem~\ref{theo:beta-convergence}]
    Convergence follows from Theorem~\ref{theo:beta-convergence-weak}.
    To prove that the bound is asymptotically tight, we construct a family of graphs
    $\{G_i\}_{i\in\NN}$, with $G_0=(\{0\},\emptyset)$, such that $G_i$ is a graph of minimal order (and size) whose $\beta$-contraction converges in exactly $i$ iterations.
    The construction, detailed in Appendix~\ref{app:fibonacci}, satisfies
    \[
        |V(G_i)| = F_{i+2}
        \quad\text{and}\quad
        G_{i+1}/\beta = G_i ,
    \]
    where $F_j$ denotes the $j$-th Fibonacci number.
    
    The graphs $G_i$ are obtained by an adversarial ordering of the vertices, which forces
    each iteration of $\beta$-contraction to reduce the graph as little as possible.
    Figure~\ref{fig:constructiveG-main} illustrates the structure of the first instances of this family.
    
    Under the well-known property that $j = \lfloor \log_\varphi(\sqrt5F_j)\rceil$, it follows that $\lfloor \log_\varphi(n) \rfloor$ iterations are necessary in the worst case, therefore concluding the proof.
\end{proof}

In light of the described results, it is then natural to claim the following two Remarks.

\begin{remark}[On the sharness of Theorem~\ref{theo:beta-convergence}]
	The esteem of $\lfloor \log_\varphi(n)\rfloor$ maximum number of iterations from Theorem~\ref{theo:beta-convergence} is asymptotically tight but not stiff.
	However, as far as $G = (V, E, \gamma_\textit{id})$ is minimally connected (\ie $|E| = m = n-1 = |V|-1$), the maximum number of iterations $i_\textit{max}$ for a given number of nodes $n$ is bounded by
   	\begin{equation}
		\lfloor \log_\varphi(n)\rfloor - 1 \leq i_\textit{max} \leq \lfloor \log_\varphi(n)\rfloor - 1\ .
	\end{equation}
	While not being surprising, since $\lfloor\log_\varphi(F_{j}-1)\rfloor$ might share the same value as $\lfloor\log_\varphi(F_{j})\rfloor$, however, the exact esteem (which is out of the scope of this work) might be derived from Binet's formula, \ie $F_j = \frac{\varphi^n-(-\varphi)^{-n}}{\sqrt{5}}$.
\end{remark}

\begin{remark}[On the role of vertex enumeration]\label{rm:random-order}
    The worst-case behaviour exhibited in Theorem~\ref{theo:beta-convergence} relies on highly structured and adversarial vertex enumerations.
    In contrast, empirical evidence suggests that for random vertex orderings the $\beta$-contraction typically converges in significantly fewer iterations than the logarithmic worst-case bound.
    While a probabilistic analysis of random enumerations is outside the scope of the present work, this observation indicates that the Fibonacci-based construction captures an extreme and non-generic regime of the dynamics.
\end{remark}


\section{The Algorithmic Framework}\label{sec:algorithmic-framework}

In this section, we describe an explicit algorithmic realisation of the $\beta$-contraction process introduced and analysed in the previous sections.
The goal is to translate the theoretical construction into a concrete and efficient procedure operating on finite graphs, while preserving the structural properties required for correctness and convergence.

In the following, we introduce the data structures used to represent graphs and contraction mappings, provide an overview of the algorithmic structure, detail the two main phases of the procedure, and analyse the resulting computational complexity.

\subsection{Data structures}\label{subsec:data-structures}

Throughout the algorithmic description, graphs are assumed to be finite, undirected, and vertex-coloured.
Vertices are implicitly enumerated in $\{0, \dots, n-1\}$, where $n = |V|$ denotes the current number of vertices.
This enumeration is updated at each contraction step.

\paragraph*{Graph representation}
A graph $G = (V,E,\gamma)$ is represented by an adjacency-list structure.
For each vertex $v \in V$, the list $N(v)$ stores the indices of its adjacent vertices.
This representation allows linear-time traversal of the edge set and supports efficient updates during contraction.
Vertex colours are stored in a separate array indexed by vertex identifiers.

We assume that the graph structure maintains no parallel edges and no self-loops.
These conditions are enforced by the contraction process described below.

\paragraph*{Contraction mapping}
A contraction mapping $\beta : V \to V'$ --denoted by \cMap-- is represented explicitly as an array \becomes of size $|V|$, where $\becomes[v]$ stores the identifier of the contracted vertex to which $v$ is mapped.
In addition, we maintain the inverse relation $\beta^{-1}$, represented as a collection \revBecomes of disjoint lists, each encoding a colour cluster to be merged into a single vertex.
This dual representation allows both forward mapping and efficient aggregation of vertices during contraction.

The contraction mapping is constructed so that vertices mapped to the same image share the same colour, in accordance with the colour-preserving property of $\beta$.

\paragraph*{Auxiliary structures}
Several auxiliary arrays are used during the construction and application of the contraction mapping, including:
(i) temporary parent pointers (\texttt{I}) supporting the construction of the oriented forest $D$ associated with $\beta$;
(ii) arrays storing cluster sizes (\cSize);
and (iii) boolean or bit-array markers used to identify active vertices during compaction (\texttt{a}).
All auxiliary structures are allocated with linear size in the number of vertices and edges of the current graph.

Unless otherwise stated, all data structures are assumed to support constant-time access by index.
Memory management details are omitted when they are not relevant to the logical structure of the algorithm.

\subsection{Overview of the \texorpdfstring{$\beta$}{beta}-contraction algorithm}\label{subsec:algo-overview}

We now provide a high-level algorithmic framework implementing the iterative $\beta$-contraction.
The procedure operates on the coloured graph $G = (V,E,\gamma)$ and applies successive contractions until the canonical colour contraction is reached.

The algorithm consists of a main loop, whose number of iterations is logarithmic in the number of vertices (cf.\ Section~\ref{subsec:beta-convergence}).
Each iteration is composed of:
(i) the generation of the contraction mapping $\cMap$, and 
(ii) the application of the induced graph contraction. 

In the first step, $\beta_k$ is constructed from the current graph $G_k$ using the local representative-selection rule defined in Section~\ref{ssec:beta-overview}.
This step implicitly builds the functional digraph $D_k$ and identifies the corresponding colour sub-partition.

In the second step, the graph is contracted according to $\beta_k$, yielding the quotient graph
\(
    G_{k+1} = G_k / \beta_k
\).
Vertices mapped to the same representative are merged, edges are updated accordingly, and vertex colours are preserved.

The algorithm terminates when the contraction mapping becomes trivial, \ie when no further reduction in the number of vertices is possible.
At this point, the current graph coincides with its canonical colour contraction.

A summary of the complete procedure is sketched as a tree of function calls in Figure~\ref{fig:functionTree}, using a compact, mathematically oriented notation.
The interested reader can find a full description of the algorithms in Appendix~\ref{app:algorithms}.

\begin{figure}[t]
    \scalebox{.88}{\input{img/contractionAlgorithm}}
    \medskip
    \caption[Function tree of $\beta$-contraction algorithm]{
		Description of the $\beta$-contraction algorithm, \ie an iteration of our novel $\gamma$-contraction algorithm, sketched as a function tree.
		Each leaf summarises a different phase by means of a simplified math-oriented pseudo-code.
	}
    \label{fig:functionTree}
\end{figure}

\subsection{Detailed construction of the contraction mapping \texorpdfstring{$\beta$}{beta}}
\label{ssec:beta-construction}

We now describe in detail the construction of the contraction mapping $\beta$ associated with a single iteration of the $\beta$-contraction algorithm.
Although $\beta$ is defined conceptually as the composition $\beta = \alpha \circ \hat\pi$, its effective construction is realised algorithmically through a single map, denoted by $\becomes$, which is progressively updated until it encodes $\beta$ explicitly.

The overall procedure implements the four conceptual steps introduced in Section~\ref{ssec:beta-overview}—namely the construction of the digraph $D$, the projection $\hat\pi$, the compaction map $\alpha$, and finally $\beta$—using a sequence of linear-time transformations.
In addition to $\becomes$, the procedure maintains:
(i) the inverse mapping $\revBecomes$, which represents $\hat\pi^{-1}$ and later $\beta^{-1}$, and
(ii) the array $\cSize$, storing the sizes of the colour clusters induced by $\beta$ (which also correspond to the sizes of the lists stored in \revBecomes).

Throughout this section, we assume that vertices are indexed as $V = \{0,\dots,n-1\}$, and we use the same notation for vertices and their indices.
Since contracted graphs are always reindexed contiguously, the maps $\hat\alpha : R \to \{0,\dots,n'-1\}$ and $\alpha : R \to V'$ coincide at the implementation level.

\paragraph*{Step 1: allocation and initialisation}
The data structure $\cMap$ is allocated and initialised so that $\becomes[v] = v$ for all $v \in V$.
At this stage, $\becomes$ represents the identity map and no contraction has yet been performed.
This step requires linear time and space in $n$.

\paragraph*{Step 2: construction of the functional digraph}
The map $\becomes$ is updated to encode the edges of the functional digraph $D$.
For each vertex $v$, the representative $\hat v$ is selected as the minimum element in $N_\gamma(v) \cup \{v\}$, and $\becomes[v]$ is set to $\hat v$.
This greedy procedure scans the full neighbourhood of each vertex and runs in $\OOO(m)$ time.

\paragraph*{Step 3: projection onto roots}
At this point, $\becomes$ defines a forest of directed trees, each corresponding to a colour cluster.
To make $\becomes$ represent the projection $\hat\pi : V \to R$, it suffices to update
\[
    \becomes[v] \lassign \becomes[\becomes[v]]
\]
for all $v \in V$.
Since $\becomes[v] \le v$ by construction, a single in-order traversal is sufficient to map every vertex directly to the root of its tree.
This step runs in linear time.

\paragraph*{Step 4: evaluation of cluster sizes}
Once $\hat\pi$ is available, the size of each colour cluster is computed as
\[
    \cSize[r] = |\{v \in V \mid \becomes[v] = r\}|
    \qquad \forall r \in R .
\]
This information is required to allocate the inverse mapping $\revBecomes$ efficiently.
The computation takes $\OOO(n)$ time.

\paragraph*{Step 5: construction of the inverse projection}
Using the cluster sizes, the inverse mapping $\revBecomes \equiv \hat\pi^{-1}$ is built so that each root $r \in R$ stores the list of vertices in its fibre.
The procedure consists of a single traversal of $\becomes$ and requires linear time and space.

\paragraph*{Step 6: compaction of clusters}
The roots are compacted into a contiguous index set $\{0,\dots,n'-1\}$ by applying the map $\alpha$.
Operationally, this is achieved by rearranging $\revBecomes$ and $\cSize$ into compact arrays, yielding a representation of $\beta^{-1}$.
The number of vertices is updated accordingly, and the operation takes $\OOO(n)$ time.

\paragraph*{Step 7: final construction of \texorpdfstring{$\beta$}{beta}}
Finally, $\becomes$ is updated so that it explicitly represents the contraction mapping $\beta : V \to V'$.
Each vertex is assigned the compact index of its colour cluster by traversing $\revBecomes$ once.
This step is linear in $n$.

\medskip

At the end of this procedure, the map $\becomes$ encodes the contraction mapping $\beta$, while $\revBecomes$ and $\cSize$ provide explicit access to the fibres and their sizes.
All steps are performed in linear time with respect to the size of the current graph, and no auxiliary data structures exceeding linear space are required.

\subsection{Application of the contraction mapping}
\label{ssec:beta-application}

We now describe the application of the contraction mapping $\beta$ to the graph structure, yielding the quotient graph
\(
    G' = G / \beta .
\)
As already anticipated in Section~\ref{ssec:beta-overview}, the contraction is performed \emph{concurrently}, meaning that all colour clusters induced by $\beta$ are merged simultaneously, rather than being contracted one at a time.
This approach avoids intermediate graph constructions and ensures that the resulting graph coincides directly with the quotient $G/\beta$.

At the algorithmic level, the application of $\beta$ decomposes into three conceptual steps: updating edge destinations, merging adjacency lists, and contracting vertex attributes.
Each step operates in place on the current graph representation and preserves the colour structure.

\paragraph*{Step 1: update edges destinations}
The contraction mapping $\beta : V \to V'$ is first applied to the endpoints of all edges.
Concretely, each occurrence of a vertex identifier in the adjacency lists is replaced by its image under $\beta$.
This operation requires a single traversal of the edge set and runs in $\OOO(m)$ time.

At this stage, adjacency lists are temporarily inconsistent, as sources still belong to $V$ while destinations already lie in $V'$.
This inconsistency is resolved in the subsequent merging phase.

\paragraph*{Step 2: merging of adjacency lists}
Vertices belonging to the same colour cluster are merged into a single vertex of the contracted graph.
For each cluster $S \in \SSS_\beta$, a new adjacency list is constructed by aggregating the adjacency lists of all vertices in $S$, after destination update.

Operationally, this corresponds to building, for each $u' \in V'$, the set
\[
    N_{G'}(u') =
    \bigcup_{v \in \beta^{-1}(u')}
    \bigl( \beta(N_G(v)) \setminus \{u'\} \bigr) .
\]
Duplicate edges created either by the contraction process or already present in the original graph are removed during this aggregation.
Self-loops are discarded, ensuring that the resulting graph remains simple.

This phase dominates the computational cost of a single contraction step.
In the worst case, aggregating neighbourhoods requires $\OOO(m)$ operations, while reconstructing adjacency lists from intermediate dense representations contributes an additional $\OOO(n'^2)$ term, where $n' = |V'|$.
Overall, this step runs in $\OOO(m + n'^2)$ time and uses $\OOO(n')$ auxiliary space.

\paragraph*{Step 3: vertex contraction}
Finally, vertex-level information is updated to reflect the contraction.
In the present setting, vertices carry only colour attributes, which are preserved under $\beta$.
Thus, a new colour array of length $n'$ is constructed by selecting, for each contracted vertex, the colour of any representative in its fibre.

This step requires linear time in the number of vertices and completes the construction of the contracted graph $G'$.

\medskip

After these three steps, the graph structure has been fully updated in place, and the resulting graph is ready to serve as input for the next iteration of the $\beta$-contraction algorithm.

\subsection{Complexity analysis}
\label{ssec:complexity-analysis}

We conclude the description of the algorithmic framework with an analysis of the computational complexity of the proposed $\beta$-contraction algorithm.
Since convergence has already been established in Section~\ref{subsec:beta-convergence}, we focus here exclusively on the cost of a single iteration and on the resulting global complexity.

\begin{table}[tp]\setlength{\tabcolsep}{6pt}
	\centering
    \footnotesize
	\begin{tabular}{r||c|c|c}
		\toprule
		 & \multirow{2}{*}{Time complexity} & \multicolumn{2}{c}{Space complexity}\\
		 & & in place & new graph\\
		\midrule\midrule
		\contractionMappingAllocation & $\OOO(n)$ & \multicolumn{2}{c}{$\tscbM{orangeM}{\OOO(n)}$}\\
		\becomesInitialisation & $\tscbM{orangeM}{\OOO(m)}$ & \multicolumn{2}{c}{$\OOO(1)$}\\
		\becomesUpdate & $\OOO(n)$ & \multicolumn{2}{c}{$\OOO(1)$}\\
		\evaluateClusterSize & $\OOO(n)$ & \multicolumn{2}{c}{$\OOO(1)$}\\
		\extractReverseBecomesMapping & $\OOO(n)$ & \multicolumn{2}{c}{$\tscbM{orangeM}{\OOO(n)}$}\\
		\revBecomesCompacting & $\OOO(n)$ & \multicolumn{2}{c}{$\OOO(n')$}\\
		\becomesCompacting & $\OOO(n)$ & \multicolumn{2}{c}{$\OOO(1)$}\\
		\midrule
		\evaluateContractionMapping& $\OOO(n+m)$ & \multicolumn{2}{c}{$\OOO(n)$}\\
		\midrule\midrule
		\edgesDestinationUpdate & $\OOO(m)$ & \multicolumn{2}{c}{$\OOO(1)$}\\
		\colorClusterMerge & $\tscbM{redM}{\OOO(n'\,^2+m)}$ & $\tscbM{orangeM}{\OOO(n')}$ & $\tscbM{orangeM}{\OOO(n'+m')}$\\
		\vertexContraction & $\OOO(n)$ & $\OOO(1)$ & $\OOO(n')$\\
		\midrule
		\applyGraphContraction& $\OOO(n'\,^2+m)$ & $\OOO(n')$ & $\OOO(n'+m')$\\
		\midrule\midrule
		\textbf{Total} & $\OOO(n+n'\,^2+m)$ & $\OOO(n)$ & $\OOO(n+m')$\\
		\bottomrule
    \end{tabular}
    \caption[Space and time bounds for $\beta$-contraction]{
        Computational time- and space complexity bounds for the proposed $\beta$-contraction algorithm.
        Per each phase, the highest computational costs are highlighted.
	}
    \label{tab:computationalCosts}
\end{table}

\paragraph*{Per-iteration complexity}
As detailed in Sections~\ref{ssec:beta-construction} and~\ref{ssec:beta-application}, a single iteration of the algorithm consists of the two phases concerning construction and application of the contraction mapping $\beta$.

The construction of $\beta$ requires a constant number of linear scans over the vertex set and a single traversal of the edge set.
Hence, its time complexity is $\OOO(n + m)$ and its auxiliary space consumption is $\OOO(n)$.

The application of the contraction mapping is dominated by the merging of adjacency lists.
In the worst case, this step requires $\OOO(n'^2 + m)$ time, where $n' = |V'|$ denotes the number of vertices after contraction.
The quadratic term arises from the reconstruction of adjacency lists from dense intermediate representations and is unavoidable in algorithms that explicitly materialise the quotient graph.
The auxiliary space required by this phase is $\OOO(n')$, excluding the memory needed to store the contracted graph itself.

Table~\ref{tab:computationalCosts} summarises the asymptotic time and space bounds of all algorithmic phases.
Overall, the cost of a single iteration is bounded by
\[
    \OOO(n + n'\,^2 + m) \text{ time} \qquad \text{and} \qquad \OOO(n) \text{ auxiliary space}.
\]

\paragraph*{Global complexity}
By Theorem~\ref{theo:beta-convergence}, the iterative application of $\beta$-contraction converges to the canonical colour contraction in $\OOO(\log n)$ iterations.
Combining this result with the per-iteration bounds yields a worst-case time complexity of
\[
    \OOO\bigl((n^2 + m)\log n\bigr),
\]
which may occur in pathological configurations where the number of vertices decreases only marginally at each iteration: it is \eg the case of graphs where $\sfrac n2$ nodes never contract, hence keeping progressively $n' \approx n$ over iterations.

In practice, however, the algorithm is designed for settings in which colour contraction is expected to significantly reduce graph size.
When $n' \ll n$ after the first few iterations --the typical case in both random and real-world graphs-- the quadratic term becomes negligible and subsequent iterations are substantially cheaper.
As a result, the observed running time is usually dominated by the first contraction step.

Empirical evidence supporting this behaviour can be found in the literature, including benchmarks on random Erd\H{o}s--R\'enyi graphs~\cite{Lombardi_Onofri_22A}, real-world social networks~\cite{Lombardi_Onofri_22}, and large-scale transaction graphs~\cite{Caprolu_Di-Pietro_Lombardi_etal_24}.

\section{Conclusions}
\label{sec:conclusions}

In this work, we introduced a rigorous and self-contained framework for colour-based graph contraction.
Starting from a formal definition of $\gamma$-contraction, we characterised graph reduction as a quotient-like operation constrained by both connectivity and vertex colours, thereby providing a precise mathematical foundation for a class of contraction procedures that are used in practice but, to the best of our knowledge, rarely formalised.
From a broader combinatorial perspective, the proposed framework complements classical contraction-based structural theories by introducing categorical information as a first-class invariant.

To address the algorithmic limitations of a fully global construction, we proposed the notion of $\beta$-contraction, a locally defined and colour-preserving contraction rule.
We showed that iterative $\beta$-contraction converges to the canonical $\gamma$-contraction, and we established structural and convergence guarantees for this process.
In particular, we proved that the number of iterations required is logarithmic in the size of the graph, with a worst-case behaviour arising from self-similar configurations, yielding the golden ratio as the sharpest logarithm basis.

Building on the theoretical framework, we presented an explicit algorithmic realisation of $\beta$-contraction.
The proposed algorithm operates in linear time for the construction of the contraction mapping and supports an efficient application of the induced quotient operation.
A detailed complexity analysis highlighted both worst-case bounds and typical behaviour, clarifying the conditions under which the method is practically effective.

Several directions for future work naturally follow from the present study.
On the algorithmic side, the local nature of $\beta$-contraction makes it well-suited for parallel and distributed implementations, which we plan to investigate formally.
From an applied perspective, further experimental evaluation on large-scale networks will help assess the practical benefits and limitations of colour-based contraction, particularly in domains such as transaction and interaction graphs.
Finally, additional combinatorial and probabilistic analyses may provide sharper average-case complexity bounds and a deeper understanding of the structures leading to worst-case behaviour.

\section*{Acknowledgments}
\noindent
E.\ Onofri is member of the ``Gruppo Nazionale Calcolo Scientifico -- Istituto Nazionale di Alta Matematica'' (GNCS--INdAM).\\
E.\ Onofri would like to extend his deepest gratitude to Dr.\ Flavio Lombardi (IAC--CNR) for his invaluable support.\\
Part of this work was fulfilled under the auspices of Roma Tre University, as part of the author's PhD program.

\bibliographystyle{elsarticle-num} 

\bibliography{arXiv_biblio.bib}

\appendix

\section{Formal proof of Convergence of the \texorpdfstring{$\beta$}{beta}-contraction of a colour component}
\label{app:fibonacci}

We have claimed in Theorem~\ref{theo:beta-convergence} that, in the worst case, $\beta$-contraction requires $\lfloor \log_\varphi(n) \rfloor$ iterations.
In order to prove this we proceed constructively.
We build a sequence of graphs $\{G_i\}_{i\in\NN}$ such that $G_i$ is a graph of minimal order (and size) whose $\beta$-contraction converges to $G_0=({0},\emptyset)$ in exactly $i$ iterations.
The construction satisfies $G{i+1}/\beta = G_i$ for all $i\in\NN$, and we show that $|V_i| = F_{i+2}$, where ${F_j}$ denotes the $j$-th Fibonacci number, defined by
\begin{equation}
    \begin{cases}
        F_{i+1} = F_{i} + F_{i-1}, & i>0\\
        F_1 = 1\\
        F_0 = 0
    \end{cases}\ .
\end{equation}

As usual, let $D_i$ be the forest generated according to Theorem~\ref{theo:D-forest}, $R_i$ its set of roots, and $\beta_i$ the associated contraction function, where we recall from \eqref{eq:betadef} that $\beta_i = \pi_i \circ \alpha_i$.
We define $Q_i = V_i\backslash R_i$ the set of non-roots in $D_i$ and we denote with $n_i$, $n_i^R$, and $n_i^Q$ the size of $V_i$, $R_i$, and $Q_i$ respectively.
Furthermore, let us also denote with $n_i^1$, and $n_i^2$ respectively the number of clusters of size 1 and 2.
Clearly,
\begin{align}
    n_i = n_i^R + n_i^Q = n_i^1 + 2n_i^2\ , \label{eq:xx1}\\
    n_i^R = n_{i-1}\ , \label{eq:xx2}\\
    n_i^R = n_i^1+n_i^2\ , \label{eq:xx3}\\
    n_i^Q = n_i^2 \label{eq:xx4}\ ,
\end{align}
where \eqref{eq:xx1} and \eqref{eq:xx2} hold by definition, \eqref{eq:xx3} holds since there is one root per cluster, and \eqref{eq:xx4} can be derived by comparing \eqref{eq:xx1} and \eqref{eq:xx3}.

We divide the remainder of the proof into four parts, for the reader's convenience.

\subsection{Proving \texorpdfstring{$n_i \geq F_{i+2}$}{ni is grater or equal to Fib(i+2)}}
We know from Lemma~\ref{lemma:charSingleTrees} that any isolated vertex in $D_i$, \ie the $n_i^1$ roots of clusters of size 1, are required to be the result of some contraction in $G_{i+1}$, hence they must be the result of some contraction of clusters of size 2 from $D_{i+1}$.
Concurrently, according to Lemma~\ref{lemma:charR}, at least one node within each of the $n_i^2$ clusters of size 2 must derive from a cluster of size 2 in $G_{i+1}$; otherwise, we would have two adjacent roots in $D_{i+1}$.
This bounds $n_{i+1}^2$ in the interval
\begin{equation}\label{eq:xx5}
    n_i^1+n_i^2 \leq n_{i+1}^2 \leq n_i^1+2n_i^2 = n_i\ .
\end{equation}
According \eqref{eq:xx3} and \eqref{eq:xx1}, the lower and upper bounds in \eqref{eq:xx5} can be written respectively as 
\begin{equation}\label{eq:xx6}
    n_i^R \leq n_{i+1}^2 \leq n_i^R+n_i^Q\ ,
\end{equation}
that is $n_{i+1}^1 \leq n_i^Q$ and $n_{i+1}^2 \geq n_i^R$.
Hence, in order to minimise $n_{i+1}$ (and thus construct a worst-case instance) or, equivalently, $n_{i+1}^2$, we assume 
\begin{equation}\label{eq:xx7}
    n_{i+1}^1 = n_i^Q\ ,
    \qquad
    n_{i+1}^2 = n_i^R\ ,
\end{equation}
and we will prove they are admissible values in the next steps.

By collecting all the above equalities, it follows that
\begin{align}
    n_{i+1}^R
    \eqop_{\eqref{eq:xx3}}
    n_{i+1}^1 + n_{i+1}^2
    \eqop_{\eqref{eq:xx7}}
    n_i^Q + n_i^R
    \eqop_{\eqref{eq:xx4}}
    n_i^2 + n_i^R
    \eqop_{\eqref{eq:xx7}}
    n_{i-1}^R + n_i^R\ ,\label{eq:xx8}\\
    n_{i+1}^Q
    \eqop_{\eqref{eq:xx4}}
    n_{i+1}^2
    \eqop_{\eqref{eq:xx7}}
    n_i^R\ ,\label{eq:xx9}\\
    n_{i+1}
    \eqop_{\eqref{eq:xx1}}
    n_{i+1}^R + n_{i+1}^Q
    \eqop_{\eqref{eq:xx9}}
    n_{i+1}^R + n_i^R
    \eqop_{\eqref{eq:xx2}}
    n_{i} + n_{i-1}\ \label{eq:xx10}.
\end{align}
It is then easy to see that we can equip \eqref{eq:xx8}, \eqref{eq:xx9}, and \eqref{eq:xx10} with proper initial conditions (they will actually be derived in the next step of the proof) in order to obtain the following systems:
\begin{equation}\label{eq:xx11}
    \begin{cases}
        n_{i+1} = n_{i} + n_{i-1}\\
        n_0 = 1\\
        n_1 = 2
    \end{cases}
    \ ,\quad
    \begin{cases}
        n_{i+1}^R = n_{i}^R + n_{i-1}^R\\
        n_0^R = 1\\
        n_1^R = 1
    \end{cases}
    \ ,\quad
    \begin{cases}
        n_{i+1}^Q = n_{i}^Q + n_{i-1}^Q\\
        n_0^Q = 0\\
        n_1^Q = 1
    \end{cases}
    \ ,
\end{equation}
\ie $n_{i} = F_{i+2}$, $n_{i}^R = F_{i+1}$, and $n_{i}^Q = F_i$.
For completeness, we obtain from \eqref{eq:xx4} and \eqref{eq:xx7} that $n_{i}^2 = F_i$ and $n_{i}^1 = F_{i+1}$ respectively.

Finally, since we have assumed \eqref{eq:xx7}, and considering that any other choice of $(n_{i+1}^1, n_{i+1}^2)$ under \eqref{eq:xx6} would result in a higher amount of nodes $n_{i+1}$ (cf.\ \eqref{eq:xx1}), it follows that $n_i \geq F_{i+2}$, hence concluding this first part of the proof.

\begin{figure}[p]
    \centering
    \includegraphics[width=.71\linewidth]{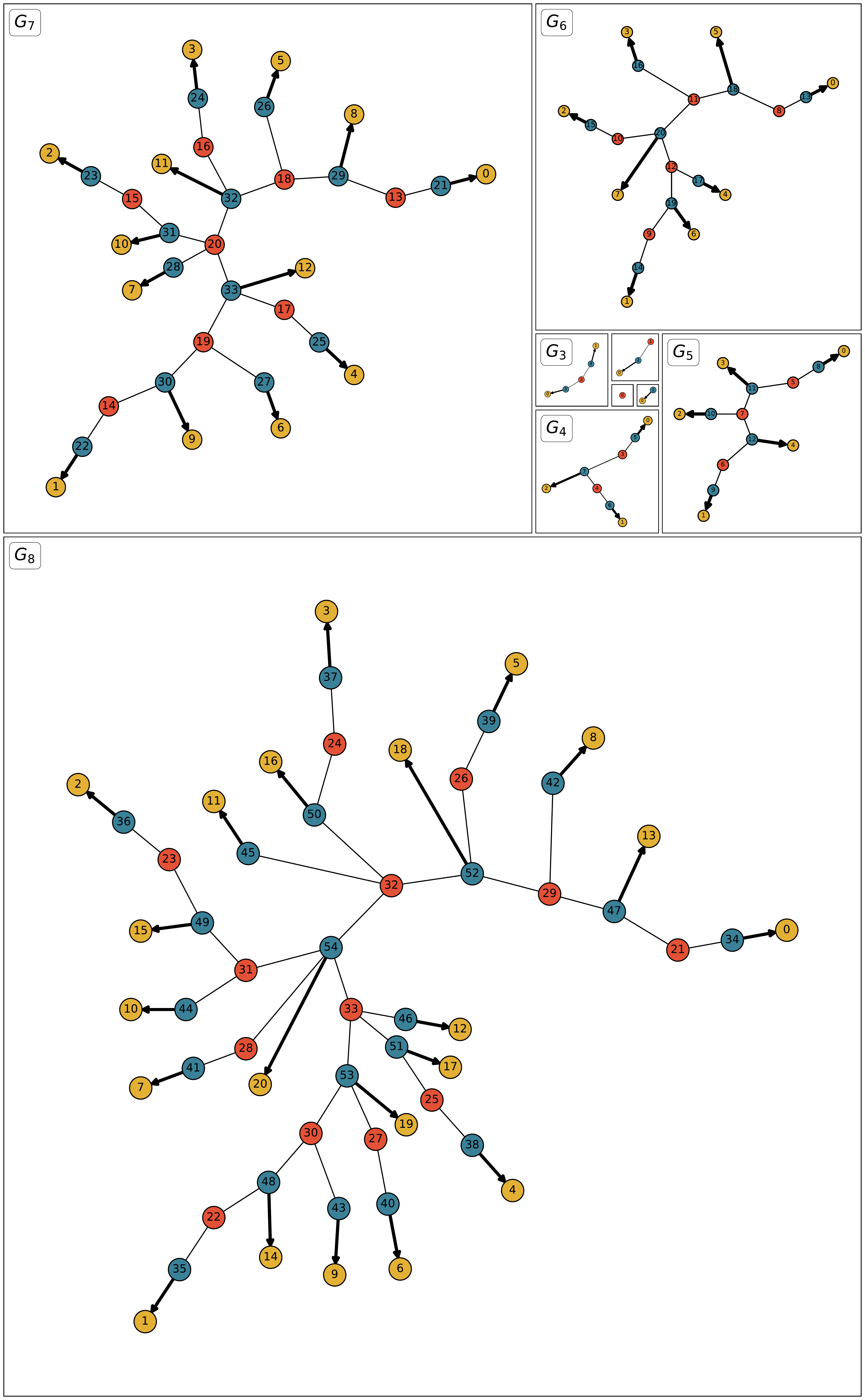}
    \\\medskip
    
    \centering
    \begin{tikzpicture}[vertex/.style={draw, circle, black}]
        \node[vertex, fill=R_P_color, label={[black,align=left]right:$R\backslash P$}] () at (6, 0) {};
        \node[vertex, fill=Q_color, label={[black,align=left]right:$Q$}] () at (3, 0) {};
        \node[vertex, fill=P_color, label={[black,align=left]right:$P$}] () at (0, 0) {};
    \end{tikzpicture}
    \caption{
        Constructive worst-case instances for $\beta$-contraction.
        The figure illustrates the sequence of graphs $\{G_i\}_{i=0}^{8}$ used in the proof of Theorem~\ref{theo:beta-convergence}.
        Vertices are coloured according to their role in the functional digraph $D_i$:
        newly added leaves $P_i$, non-leaf roots $R_i \setminus P_i$, and non-root vertices $Q_i$.
        At each step, the construction enforces $G_{i+1}/\beta = G_i$ while minimising $|V_i|$, yielding Fibonacci growth and logarithmic convergence.
    }
    \label{fig:constructiveG}
\end{figure}

\subsection{Constructive definition of minimal \texorpdfstring{$G_i$}{G(i)}}
The previous part of this proof provided us with the bound \eqref{eq:xx6} on $n_i$, depending on $i$.
We can now prove constructively that the choice in \eqref{eq:xx7} is admissible by building the sequence of graphs $G_i$ according to the previously mentioned properties (cf.\ Figure~\ref{fig:constructiveG}).

Given $G_i$ (starting with $G_0$) we define $G_{i+1}$ by adding one node $v$ per each root $r \in D_i$ and making it adjacent to $r$.
Let $P_{i+1}$ denote the set of such newly added vertices.
The idea is then to change the enumeration to make $r \in D_i$ collapse on the corresponding $v \in P_{i+1}$, \ie make $v$ a root in $D_{i+1}$ such that $r$ (its only neighbour) will become part of the same cluster of size two.
This re-enumeration is admissible since $\beta$ depends on vertex ordering.
At the same time, non-root elements in $D_i$, \ie elements from $Q_{i}$, should become roots in $D_{i+1}$;
this way we would obtain
\begin{equation}\label{eq:xx12}
    R_{i+1} = Q_i \cup P_{i+1} \qquad \mbox{and} \qquad Q_{i+1} = R_i\ ,
\end{equation}
hence obtaining \eqref{eq:xx7} since $|P_{i+1}| = |R_i|$.

In order to achieve this result, we change the enumeration of the vertices in $V_{i+1}$ as follows:
\begin{itemize}
    \item $\forall r\in R_i$ enumerated as $j$ in $G_i$, we enumerate the novel vertex attached to it as $j$ and we enumerate $r$ as $n_{i}+j$.
    
    \item $\forall v \in Q_i$, we keep their enumeration as in $G_i$.
\end{itemize}

Clearly, we have
\begin{equation}\label{eq:rel_u_u*}
    u^*\geq u\ \quad \forall u \in V_i, \qquad \mbox{and} \qquad \text{equality holds }\iff u \in Q_i 
\end{equation}
where $u^* \in V_{i+1}$ is its counterpart in $G_{i+1}$.

We now prove \eqref{eq:xx12} in three points, analysing $P_{i+1}$, $R_{i}$, and $Q_{i}$ in order:
\begin{enumerate}
    \item It is straightforward to see that each novel vertex $v_j \in P_{i+1}$ is connected with $v_{n_i+j} \in V_{i+1}$ only, and hence $v_j \in R_{i+1}$, \ie $\pi_{i+1}(v_j) = v_j$ and $P_{i+1} \subseteq R_{i+1}$.

    \item Concurrently, $v_j \in P_{i+1}$ is the smallest amongst the nodes in $N(v_{n_i+j})$: in fact $v_{j+n_i}$ was enumerated as $j$ in $G_i$ and it was a root in $D_i$, hence $v_j$ is smaller than any of its other neighbours in $G_i$, and the same holds for $G_{i+1}$ according to \eqref{eq:rel_u_u*}; consequently $v_{n_i+j} \in Q_{i+1}$, $\pi(v_{n_i+j}) = v_j$, and $R_i \subseteq Q_{i+1}$.

    \item Finally, it is easy to see by induction that vertices in $Q_{i}$ are pairwise disjoint and form roots in $D_{i+1}$, with the base of induction being trivially true since $Q_0 = \emptyset$.
        Then, assuming vertices in $Q_{i-1}$ disjoint, it follows that $\forall v_l \in Q_{i-1}$ we have $N(v_l) \subseteq R_{i-1}$ \ie nodes $r_j \in N(v_l)$ are all roots that at iteration $i$ get enumerated with indices $j+n_{i-1}>l$ and, consequently, $v_l \in R_{i}$, $\pi_i(v_l) = v_l$, and $Q_{i-1} \subseteq R_{i}$.
        Consequently, it follows from the previous point that $Q_i = R_{i-1}$, hence proving the inductive hypothesis.
        At the same time, we have that $R_i = P_i \cup Q_{i-1}$, hence concluding the proof for \eqref{eq:xx12}.
\end{enumerate}
Again, it follows by construction that $|P_i| = |R_{i-1}|$, and hence \eqref{eq:xx7} holds for our construction.
Furthermore, it is easy to see (also pictorially, cf.\ Figure~\ref{fig:constructiveG}) that initial conditions for \eqref{eq:xx11} are verified under this construction. We can now proceed to prove that $G_{i+1}/\beta = G_i$ so that $n_i = F_{i+2}$ holds.

\subsection{Proving the equality between \texorpdfstring{$G_{i+1}/\beta$}{G(i+1)/beta} and \texorpdfstring{$G_{i}$}{G(i)}} 

In order to simplify the proof of $G_{i+1}/\beta =_{(?)} G_i$, we first proceed by describing the ordering of the nodes within $Q_i$, $R_i$, and $P_i$.

In particular, it is easy to see that
\begin{align}
    R_i = \{0, \dots, n_{i}^R-1\}\ = \{0, \dots, n_{i-1}-1\}\ ,\label{eq:xx13}\\
    P_i = \{0, \dots, n_{i-1}^R-1\} \subseteq R_i\ ,\label{eq:xx14}\\
    Q_i = \{n_i^R, \dots, n_{i}-1\}\ = \{n_{i-1}, \dots, n_{i-1}+n_{i-1}^R-1\}\ .\label{eq:xx15}
\end{align}
In particular, \eqref{eq:xx15} follows by the fact that $Q_i = R_{i-1}$ and roots in $D_{i-1}$ are enumerated as $n_{i-1}+j$ in $V_i$.
Consequently, \eqref{eq:xx13} holds true since $R_i = V_i\backslash Q_i$.
Finally, \eqref{eq:xx14} follows since $P_i$ are enumerated as $R_{i-1} = \{0, \dots, n_{i-1}-1\}$ (and $R_0 = \{0\}$).

Combining the novel enumeration with the results of (1)--(3) from the previous part of the proof, we can now state that:
\begin{equation}\label{eq:pi-beta-proof}
    \pi_{i+1}(v_j) = \begin{cases}
        v_j & \mbox{if } j < n_i\\
        v_{j-n_i} & \mbox{otherwise}
    \end{cases}
\end{equation}
In particular, since $R_{i+1}$ is already compacted in $\{0, n_{i+1}^R\}$, we have that $\alpha_{i+1} = id$, \ie the identity function. We can conclude that
\begin{equation}
    \beta_{i+1}(v_j) = \begin{cases}
        v_j & \mbox{if } j < n_i\\
        v_{j-n_i} & \mbox{otherwise}
    \end{cases}\ ,
\end{equation}
where we recall that $\beta:V_{i+1}\to V_i$.
Hence, we can conclude this part of the proof since it is then trivial that $G_{i+1}/\beta = G_i$, given $G_{i+1}$ built from $G_{i}$, which also proves that the bound from the first part of the proof is indeed an equality, \ie
\begin{equation}\label{eq:fibonacciNodes}
    n_i = F_{i+2}\ .
\end{equation}

\subsection{Proving \texorpdfstring{$\log_\varphi(n)$}{log(varphi, n)} is an upper bound for the convergence time}
We proved in the previous step that the constructed $G_i$, for $i \in \NN$, are such that $G_i$ is a minimal order graph that takes $i$ iterations to converge to $G_0=(\{0\}, \emptyset)$ and that $n_i = F_{i+2}$, see~\eqref{eq:fibonacciNodes}.

We recall the following well-known property for the Fibonacci sequence:
\begin{equation}
    j = \lfloor \log_\varphi(\sqrt{5}F_j)\rceil\ ,
\end{equation}
with the rounding being inherently small, and in particular $<0.1$ for $i\geq4$.
Hence, we obtain
\begin{equation}
    j = \lfloor \log_\varphi(\sqrt{5}) + \log_\varphi(F_j)\rceil = \lfloor 1.67 + \log_\varphi(F_j) + \epsilon\rceil = 2 + \lfloor \log_\varphi(F_j)\rfloor \ ,
\end{equation}
where $\epsilon < 0.1$ and the last equality is derived by the fact that the total error is given by $\epsilon$ plus rounding error, and, consequently, is bounded by $0.2$.

It follows that, being $n_i = F_{i+2}$, then the number of iterations can be derived as $i = \lfloor(\log_\varphi(n_i))\rfloor$, hence concluding the proof due to the minimality of $G_i$.


\section{Implementation of the \texorpdfstring{$\beta$}{beta}-Contraction Algorithm}
\label{app:algorithms}

This appendix collects a complete reference implementation of the $\beta$-contraction algorithm described in Section~\ref{sec:algorithmic-framework}, a working version of which is available online at \url{https://github.com/eOnofri04/GraphColorContraction}.
The material is intended for readers interested in implementation details, reproducibility, or practical deployment, and can be safely skipped without loss of continuity.
All theoretical notions required --including contraction mappings, functional digraphs, and colour clusters-- are assumed from the main body of the work.

\subsection{Data structures}
\label{app:data-structures}

Our reference implementation is written in \Clang{} and adopts a minimal, array-based representation.
Vertices are indexed contiguously, and all mappings are stored explicitly as arrays to ensure predictable memory access patterns.

Throughout this appendix, we use \texttt{typewriter font} for code-level entities.
Unless otherwise stated, vertex indices are stored using an unsigned integer type, denoted by \idxt, whose concrete width depends on the target platform.
Vertex colours are stored using a dedicated type \colt.

Arrays (denoted by prepending $\ptr$ characters) are explicitly allocated and deallocated using standard C-style memory management to properly delimit their lifespan; allocation details are omitted when not relevant to the algorithmic logic.

\begingroup
\begin{algorithm}[tp]
    \caption{\Grapht data structure}
    \label{algo:graphDataStructure}
        \KwStruct{\Grapht}
        \begin{algorithmic}[1]
            \State \idxt $n$;          \Comment{order of the graph}
            \State \idxt $m$;          \Comment{size of the graph}
            \State \colt \ptr\Ctt;     \Comment{colour list, of length $n$}
            \State \idxt \ptr\Nvtt;    \Comment{vertices degree, of length $n$}
            \State \idxt \ptr\ptr\Ett; \Comment{adjacency lists, for each vertex $v$ of length \Nvtt[$v$]}
        \end{algorithmic}
\end{algorithm}
\begin{algorithm}[tp]
	\caption{\cMapt data structure}
	\label{algo:cMapDataStructure}
	\KwStruct{\cMapt}
            \begin{algorithmic}[1]
                
		\State \idxt $n$;                 \Comment{order of the original graph}
		\State \idxt $n'$;                \Comment{order of the contracted graph, $|\SSS_\beta|$}
		\State \idxt $n^*$;               \Comment{effective size of $\cSize$ and $\revBecomes$, either $n$ or $n'$}
		\State \idxt \ptr\cSize;          \Comment{size of the colour clusters $S \in \SSS$, of length $n^*$ or $n'$}
		\State \idxt \ptr\becomes;        \Comment{$\beta$ mapping, of length $n$}
		\State \idxt \ptr\ptr\revBecomes; \Comment{$\beta^{-1}$ mapping, for each vertex $v$ of length \cSize[$v$]}
            \end{algorithmic}
\end{algorithm}
\endgroup

Graphs are represented using adjacency lists.
The graph structure \Grapht{}, defined in Algorithm~\ref{algo:graphDataStructure}, stores degrees and colours alongside the adjacency lists.
The contraction mapping $\beta$ and its inverse are stored in a structure
\cMapt{}, defined in Algorithm~\ref{algo:cMapDataStructure}; its fields \becomes{} and \revBecomes{} are progressively updated during the construction of the contraction.
The intermediate maps discussed in the theoretical sections are not individually stored
explicitly.

\subsection{Implementation of the algorithm}\label{ssec:algoInDepth}

At a high level, the algorithm takes in input a graph $G = (V, E, \gamma)$ and progressively evaluates the $\beta$-contraction operation $G' \lassign G/\beta$ in place, until $G/\gamma$ is obtained (see Algorithm~\ref{algo:graphContraction}).

\begin{algorithm}[tp]
    \caption{$\GraphColorContraction(\Gtt)\algoout\Gtt'$}
    \label{algo:graphContraction}
    \KwIn{The \Grapht $\Gtt$ to be contracted}\\
    \KwOut{The \Grapht $\Gtt'$, colour contraction of $\Gtt$}
    
    \begin{algorithmic}[1]
        \State $\cMap \lassign \evaluateContractionMapping(\Gtt)$;
        \While{$\cMap.n \ne \cMap.n'$}
            \State $\applyGraphContraction(\Gtt, \cMap)$;
            \State $\texttt{freeCMap}(\cMap)$; \Comment{Opportunely $\freefun$s allocated elements in $\cMap$}
            \State $\cMap \lassign \evaluateContractionMapping(\Gtt)$;
        \EndWhile
        \State $\texttt{freeCMap}(\cMap)$; \Comment{Opportunely $\freefun$s allocated elements in $\cMap$}
        \State \Return \Gtt;
    \end{algorithmic}
\end{algorithm}

The algorithms listed below provide a complete serial implementation of a single iteration of $\beta$-contraction and of the overall iterative procedure.
They directly reflect the algorithmic structure described in Section~\ref{subsec:algo-overview} and follow the same naming conventions.

Let $G = (V, E, \gamma)$ and $G' = (V', E', \gamma') = G/\beta$ and let $n = |G|$, $m = \|G\|$, $n' = |G'|$, and $m' = \|G'\|$. We recall that here $V = \{0, \dots, n-1\}$ and $V' = \{0, \dots, n'-1\}$, hence, when we write $\forall v \in V$, we mean that a variable \idxt $v$ is ranging in the interval $0\leq v<n$.
The pseudocode for the two phases of $\beta$-contraction is presented in Algorithm~\ref{algo:evaluateContractionMapping} (generation of the contraction mapping $\cMap$) and Algorithm~\ref{algo:applyGraphContraction} (\cMap application).

\begin{algorithm}[tp]
    \caption{$\evaluateContractionMapping(\Gtt) \algoout \cMap$}
    \label{algo:evaluateContractionMapping}
    \KwIn{The \Grapht $\Gtt$}\\
    \KwOut{The corresponding \cMapt holding $\beta$ and $\beta^{-1}$}

    \begin{algorithmic}[1]
        \State $\cMapt\ \cMap$;
        \Statex
        \State $\cMap \lassign \contractionMappingAllocation(\Gtt.n)$;
        \State $\becomesInitialisation(\Gtt, \cMap)$;
        \State $\becomesUpdate(\cMap)$;
        \State $\evaluateClusterSize(\cMap)$;
        \State $\extractReverseBecomesMapping(\cMap)$;
        \State $\revBecomesCompacting(\cMap)$;
        \State $\becomesCompacting(\cMap)$;
        \Statex
        \State\Return \cMap;
    \end{algorithmic}
	
\end{algorithm}

\begin{algorithm}[tp]
    \caption{$\applyGraphContraction(\Gtt, \cMap)$}
    \label{algo:applyGraphContraction}
    
    \KwIn{The \Grapht $\Gtt$ to be contracted with the $\beta$-map $\cMap$}\\
    \KwOut{Nothing, contraction is performed in place}
    \begin{algorithmic}[1]
        \State $\edgesDestinationUpdate(\Gtt, \cMap)$;
        \State $\colorClusterMerge(\Gtt, \cMap)$;
        \State $\vertexContraction(\Gtt, \cMap)$;
    \end{algorithmic}
\end{algorithm}

In the pseudo-codes that follow (Algorithm~\ref{algo:graphContraction}--\ref{algo:vertexContraction}), we keep naming consistency as much as possible; we here list a few variable names that are consistently used and that we are not re-defining in each pseudo-code for the sake of simplicity:
\begin{itemize}
	\item \Grapht $\ptr\Gtt$ is the graph $G=(V, E)$ as defined in Algorithm~\ref{algo:graphDataStructure}.
	\item \Grapht $\ptr\Gtt'$ is the contracted graph $G/\beta = (V', E')$. Do note that the algorithms described work in-place to avoid multiple allocations, hence $\Gtt'$ is never formally allocated.
	\item \cMapt $\ptr\cMap$ is the contraction mapping as defined in Algorithm~\ref{algo:cMapDataStructure}.
	\item \idxt $i, j, k$ are (indices of) vertices within $V$, ranging in $[0, n)$.
	\item \idxt $i', j', k'$ are (indices of) vertices within $V'$, ranging in $[0, n')$.
	\item \idxt $\iidx, \jidx, \kidx$ define (indices of) vertices within $\Ett[\cdot]$ (\ie edges), ranging in $[0, \Nvtt[\cdot])$.
	\item \idxt $\ibdx, \jbdx, \kbdx$ define (indices of) vertices within $\revBecomes[\cdot]$, ranging in $[0, \Csize[\cdot])$.
\end{itemize}

\subsubsection{Contraction mapping creation}

As highlighted in Algorithm~\ref{algo:evaluateContractionMapping}, the construction of the contraction mapping is decomposed into seven steps.

\paragraph*{Allocation and initialisation (Algorithm~\ref{algo:contractionMappingAllocation})}
The contraction map \cMap{} is allocated according to the current graph size $n$ and initialised so that
\[
    \becomes[v] = v \qquad \forall v \in V \ .
\]
Accordingly, $\cMap.n^* \lassign n$.
The inverse mapping \revBecomes{} is not allocated at this stage, as cluster sizes are still unknown.

\begin{algorithm}[tp]
    \caption{$\contractionMappingAllocation(n)\ \algoout\ \cMap$}
    \label{algo:contractionMappingAllocation}
    \begin{algorithmic}[1]
        \State $\cMapt\ \cMap$;
        \State $\cMap \lassign \alloc(1, \cMapt)$;
        \State $\cMap.n \lassign n$;
        \State $\cMap.n' \lassign 0$;
        \State $\cMap.n^* \lassign n$;
        \State $\cMap.\becomes \lassign \alloc(n, \idxt)$;
        \State $\cMap.\revBecomes \lassign \alloc(n, \idxt\ptr)$;
        \State $\cMap.\cSize \lassign \alloc(n, \idxt)$;
        \For{$i \in \{0, \dots, n-1\}$}
            \State $\cMap.\becomes[i] \lassign i$;
        \EndFor
        \State \Return \cMap;
    \end{algorithmic}
    
\end{algorithm}

	
\paragraph*{Construction of the functional digraph (Algorithm~\ref{algo:becomesInitialisation})}
The array \becomes{} is updated to encode the edge set of the functional digraph $D$ by scanning all adjacency lists and selecting, for each vertex, its local representative.
    
\begin{algorithm}[tp]
    \caption{$\becomesInitialisation(\Gtt, \cMap)$}
    \label{algo:becomesInitialisation}
    \begin{algorithmic}[1]
        \State \colt $c$; \Comment{colour of current vertex}
        \State \idxt $b$; \Comment{current best choice for becomes}
        \Statex
        \ForAll{$i \in \{0, \dots, \Gtt.n-1\}$}
            \State $c \lassign \Gtt.\Ctt[i]$;
            \State $b \lassign i$;
            \ForAll{$\jidx \in \{0, \dots, \Gtt.\Nvtt[i]-1\}$}
                \State $j \lassign \Gtt.\Ett[i][\jidx]$;
                \If{$j < b \land \Gtt.\Ctt[j] = c$}
                    \State $b \lassign j$;
                \EndIf
            \EndFor
            \State $\cMap.\becomes[i] \lassign b$;
        \EndFor
    \end{algorithmic}
\end{algorithm}


\paragraph*{Projection onto roots (Algorithm~\ref{algo:becomesUpdate})}
This step updates \becomes{} so that each vertex is mapped directly to the root of its tree in the functional digraph.
In the serial setting, a single in-order traversal suffices.
    
\begin{algorithm}[tp]
    \caption{$\becomesUpdate(\cMap)$}
    \label{algo:becomesUpdate}
    \begin{algorithmic}[1]
        \ForAll{$i \in \{1, \dots, \cMap.n-1$}
            \Comment{Must be executed in order to work correctly}
            \State $\cMap.\becomes[i] \lassign \cMap.\becomes[\cMap.\becomes[i]]$;
        \EndFor
    \end{algorithmic}
\end{algorithm}

	
\paragraph*{Evaluation of cluster sizes (Algorithm~\ref{algo:evaluateClusterSize})}
Cluster sizes are computed by a linear traversal of \becomes{}, counting the number of vertices mapped to each root.
The resulting array \cSize{} is used to preallocate the inverse mapping.
    
\begin{algorithm}[tp]
    \caption{$\evaluateClusterSize(\cMap)$}
    \label{algo:evaluateClusterSize}
    \begin{algorithmic}[1]
        \ForAll{$i \in \{0, \dots, \cMap.n-1\}$}
            \State $\cMap.\cSize[\cMap.\becomes[i]]\plusplus$;
        \EndFor
        
        \ForAll{$i \in \{0, \dots, \cMap.n-1\}$}
            \If{$\cMap.\cSize[i] > 0$}
                \State $\cMap.n'\plusplus$;
            \EndIf
        \EndFor
    \end{algorithmic}
\end{algorithm}

	
\paragraph*{Construction of the inverse projection (Algorithm~\ref{algo:extractReverseBecomesMapping})}
Using \cSize{}, the inverse map \revBecomes{} is allocated and filled by a single traversal of \becomes{}.
An auxiliary index array $\Itt{}$ tracks the current insertion position for each cluster.

\begin{algorithm}[tp]
    \caption{$\extractReverseBecomesMapping(\cMap)$}
    \label{algo:extractReverseBecomesMapping}

    \begin{algorithmic}[1]
        \State \idxt \ptr$\Itt$; \Comment{counter for $\jbdx$ of $\revBecomes[\becomes[i]]$}
        
        \State $\Itt \lassign \alloc(\cMap.n, \idxt)$;
        
        \Statex
        
        \ForAll{$i \in \{0,\dots,\cMap.n-1\}$}
            \If{$\cMap.\cSize[i] > 0$}
                \State $\cMap.\revBecomes[i] \lassign \alloc(\cMap.\cSize[i], \idxt)$;
            \Else
                \State $\cMap.\revBecomes[i] \lassign \NULL$;
            \EndIf
        \EndFor
        
        \Statex
        
        \ForAll{$i \in \{0, \dots, \cMap.n-1\}$}
            \State $i' \lassign \cMap.\becomes[i]$;
            \State $\jbdx \lassign \Itt[i']\plusplus$;
            \State $\cMap.\revBecomes[i'][\jbdx] \lassign i$;
        \EndFor
        
        \Statex
        
        \State $\freefun(\Itt)$;
    \end{algorithmic}
    
\end{algorithm}


\paragraph*{Compaction of clusters (Algorithm~\ref{algo:revBecomesCompacting})}
The compaction map $\alpha : R \to V'$ is applied by rearranging \revBecomes{} and \cSize{} into compact arrays of size $n'$.
After this step, $\revBecomes$ represents $\beta^{-1}$ and $\cMap.n^*$ is updated to $n'$.

\begin{algorithm}[tp]
    \caption{$\revBecomesCompacting(\cMap)$}
    \label{algo:revBecomesCompacting}
    \begin{algorithmic}[1]
        \State \idxt $i' \lassign 0$; \Comment{explicit assignment needed since it is used as a counter}
        \State \idxt $\ptr\ptr\revBecomes', \ptr\cSize'$; \Comment{compact version of $\revBecomes$ and $\Csize$}
        
        \Statex
        
        
        \State $\Csize' \lassign \alloc(\cMap.n', \idxt)$;
        \State $\revBecomes' \lassign \alloc(\cMap.n', \idxt \ptr)$;
        
        \Statex
        
        \ForAll{$i \in \{0, \dots, \cMap.n-1\}$}
            \Comment{$\hat\alpha$ follows the roots order in $R$ if done in order}
            \If{$\cMap.\revBecomes[i] \ne \NULL$}
                \State $\Csize'[i'] \lassign \cMap.\Csize[i]$;
                \State $\revBecomes'[i'] \lassign \cMap.\revBecomes[i]$;
                \State $i'\plusplus$;
            \EndIf
        \EndFor
        
        \Statex
        
        
        \State $\freefun(\cMap.\Csize)$;
        \State $\cMap.\Csize \lassign \Csize'$;
        
        \Statex
        
        \Statex \Comment{We only $\freefun$ $\revBecomes$ array, since $\revBecomes[\cdot]$ are still used in $\revBecomes'$}
        \State $\freefun(\cMap.\revBecomes)$;
        \State $\cMap.\revBecomes \lassign \revBecomes'$;
        
        \Statex
        
        \State $\cMap.n^* \lassign \cMap.n'$;
    
    \end{algorithmic}
    
\end{algorithm}


\paragraph*{Final construction of $\beta$ (Algorithm~\ref{algo:becomesCompacting})}
The array \becomes{} is updated to explicitly encode $\beta : V \to V'$ by traversing \revBecomes{} once.

\begin{algorithm}[tp]
    \caption{$\becomesCompacting(\cMap)$}
    \label{algo:becomesCompacting}
    
    
    \begin{algorithmic}[1]
        \ForAll{$i' \in \{0, \dots, \cMap.n'-1\}$}
            \ForAll{$\jidx \in \{0, \dots, \cMap.\Csize[i']-1\}$}
                \State $\cMap.\becomes[\cMap.\revBecomes[i'][\jidx]] \lassign i'$;
            \EndFor
        \EndFor
    \end{algorithmic}
\end{algorithm}

\subsubsection{Contraction mapping application}

As highlighted in Algorithm~\ref{algo:applyGraphContraction}, the application of the contraction mapping consists of three steps.


\paragraph*{Update of edge destinations (Algorithm~\ref{algo:edgesDestinationUpdate})}

The mapping $\beta$ (stored in \becomes{}) is applied to all edge endpoints in $\Ett[\cdot]$.
After this step, adjacency lists temporarily map sources in $V$ to destinations in $V'$.

\begin{algorithm}[tp]
    \caption{$\edgesDestinationUpdate(\Gtt, \cMap)$}
    \label{algo:edgesDestinationUpdate}
    \begin{algorithmic}[1]
        \ForAll{$i \in \{0, \dots, \Gtt.n-1\}$}
            \ForAll{$\jidx \in \{0, \dots, \Gtt.\Nvtt[i]\}$}
                \State$\Gtt.\Ett[i][\jidx] \lassign \cMap.\becomes[\Gtt.\Ett[i][\jidx]]$;
            \EndFor
        \EndFor
    \end{algorithmic}
\end{algorithm}

	
\paragraph*{Merging of adjacency lists (Algorithm~\ref{algo:colorClusterMerge})}

For each colour cluster $S \in \SSS_\beta$, the adjacency list of the contracted vertex $u' = \tilde\alpha(S)$ is constructed by aggregating the updated adjacency lists of vertices in $S$.
A dense bit-array $\att$ of length $n'$ is used to evaluate the aggregation by means of logical operations and implicitly eliminate duplicate edges and self-loops.
The bit-array should be adequately substituted if more complex merge operations are required, like in the case of edge-weighted graphs: as an example, we can keep track of how many edges have been merged together by implementing a suitable function here.

\begin{algorithm}[tp]
    \caption{$\colorClusterMerge(\Gtt, \cMap)$}
    \label{algo:colorClusterMerge}
    \begin{algorithmic}[1]
        \State \idxt $\aidx$; \Comment{index of $\att$}
        \State \idxt $d$; \Comment{temporary value for $\Nvtt'[i']$}
        \State \idxt $\ptr\ett$; \Comment{temporary reference to $\Ett'[i']$, \ie sparse version of $\att$}
        \State \boolt $\ptr\att$; \Comment{temporary dense $n'$-length bit-array for the adjacency list $\ett$}
        \State \idxt $m' \lassign 0$; \Comment{value $m$ for $G'$}
        \State \idxt $\ptr\Nvtt'$; \Comment{neighbourhood degrees $\Nvtt$ for $G'$, of size $n'$}
        \State \idxt $\ptr\ptr\Ett'$; \Comment{adjacency lists $\Ett$ for $G'$, of size $n'$}
        
        \Statex
        
        \State $\Nvtt' \lassign \alloc(\cMap.n', \idxt)$;
        \State $\Ett'  \lassign \alloc(\cMap.n', \idxt \ptr)$;
        \State $\att   \lassign \alloc(\cMap.n', \booltt)$;
        
        \Statex
        
        \ForAll{$i' \in \{0, \dots, \cMap.n'-1\}$}
            \Statex{\hspace{1em} $\triangleright$ Reset $\att$}
            \ForAll{$\aidx \in \{0, \dots, \cMap.n'-1\}$}
                \State $\att[\aidx] \lassign \falsett$;
            \EndFor
            \State $d \lassign 0$;

            \Statex
            \Statex {\hspace{1em} $\triangleright$ fill $\att$ and deallocate old adjacency vectors}
            \ForAll{$\jbdx \in \{0, \dots, \cMap.\Csize[i']-1$}
                \Comment{{cycle over $\revBecomes[i']$ vertices j}}
                \State $j \lassign \cMap.\revBecomes[i'][\jbdx]$;
                \ForAll{$\kidx \in \{0, \dots, \Gtt.\Nvtt[j]-1\}$}
                    \Comment{cycle in the neighbourhood of $j$}
                    \State $k' \lassign \Gtt.\Ett[j][\kidx]$;
                    \If{$i' \ne k' \land \lnot \att[k']$}
                        \Comment{We can drop $i' \ne k'$ if self loops are allowed}
                        \State $\att[k'] \lassign \truett$;
                        \State $d\plusplus$;
                    \EndIf
                \EndFor
                \State $\freefun(\Gtt.\Ett[j])$;
            \EndFor
            
            \Statex
        
            \Statex{\hspace{1em} $\triangleright$ convert from bit-array $\att$ to adjacency list $\ett$}
            \State $\ett \lassign \alloc(d, \idxt)$;
            \State $\kidx \lassign 0$;
            \ForAll{$\aidx \in \{0, \dots, \cMap.n'-1\}$}
                \If{$\att[\aidx]$}
                    \State $\ett[\kidx\plusplus] \lassign \aidx$;
                \EndIf
            \EndFor

            \Statex
        
            \State $\Nvtt'[i'] \lassign d$;
            \State $\Ett[i'] \lassign \ett$;
            \State $m' \lassign m' + d$;
            
        \EndFor
        
        \Statex
        
        \State$\freefun(\att)$;
        \State $\freefun(\Gtt.\Nvtt)$;
        \qquad $\Gtt.\Nvtt \lassign \Nvtt'$;
        \State $\freefun(\Gtt.\Ett)$;
        \qquad $\Gtt.\Ett \lassign \Ett'$;
        \State $\Gtt.m \lassign m'$;
        
    \end{algorithmic}
\end{algorithm}


\paragraph*{Vertex contraction (Algorithm~\ref{algo:vertexContraction})}

Vertex attributes are updated to reflect the contraction.
Since vertices carry only colour information, a new colour array of length $n'$ is constructed by selecting one representative per cluster.
A merge function for vertices should be properly defined if more complex graphs are considered, like in the case of vertex-weighted graphs: as an example, we can keep track of how many nodes or edges have been merged in a vertex by implementing a suitable function here.

\begin{algorithm}[tp]
    \caption{$\vertexContraction(\Gtt,\cMap)$}
    \label{algo:vertexContraction}
    \begin{algorithmic}[1]
        \State \colt $\ptr\Ctt'$; \Comment{Colour array $\Ctt$ for $G'$, of size $n'$}
        
        \State $\Ctt' \lassign \alloc(\cMap.n', \colt)$;
        
        \Statex
        
        \ForAll{$i' \in \{0, \dots, \cMap.n'-1\}$}
            \State $\Ctt'[i'] \lassign \Gtt.\Ctt[\cMap.\revBecomes[i'][0]]$;
        \EndFor
        
        \Statex
        
        \State $\freefun(\Gtt.\Ctt)$;
        \State $\Gtt.\Ctt\lassign \Ctt'$;
        
        \State $\Gtt.n \lassign \cMap.n'$;
        
    \end{algorithmic}
\end{algorithm}

\end{document}